\documentclass[a4paper]{article}
\usepackage{amsmath}
\usepackage{amssymb}
\usepackage[english]{babel}
\usepackage[utf8]{inputenc}
\usepackage{tasks}
\usepackage[version=4]{mhchem}
\usepackage[T1]{fontenc}
\usepackage{textcomp,gensymb}

\usepackage[toc,page]{appendix}
\usepackage[a4paper,top=3cm,bottom=2cm,left=3cm,right=3cm,marginparwidth=1.75cm]{geometry}
\usepackage{cite}

\usepackage{xcolor}
\usepackage{soul}
\usepackage{subcaption}
\usepackage{amsmath}
\usepackage{color}
\usepackage{float}
\usepackage{graphicx}
\usepackage[colorinlistoftodos]{todonotes}
\usepackage[colorlinks=true, allcolors=blue]{hyperref}
\usepackage{subcaption} 
\usepackage[affil-it]{authblk}
\usepackage{array}
\newcolumntype{P}[1]{>{\centering\arraybackslash}p{#1}}

\begin{document}
\title{Machine-learning-enhanced time-of-flight mass spectrometry analysis}

\author[1]{Ye Wei*}
\author[1]{Rama Srinivas Varanasi}
\author[1]{Torsten Schwarz}
\author[1]{Leonie Gomell}
\author[1]{Huan Zhao}
\author[4]{David J. Larson}
\author[1]{Binhan Sun}
\author[3]{Geng Liu}
\author[3]{Hao Chen}
\author[1]{Dierk Raabe}
\author[1, 2]{Baptiste Gault*}
\affil[1]{Max-Planck-Institut für Eisenforschung, Max-Planck-Strasse 1, 40237 D\"usseldorf, Germany}
\affil[2] {Department of Materials, Royal School of Mines, Imperial College, London, SW7 2AZ, UK}
\affil[3] {Key Laboratory for Advanced Materials of Ministry of Education, School of Materials Science and Engineering, Tsinghua University, Beijing 100084, China}
\affil[4] {CAMECA Instruments, Inc, 5470 Nobel Drive, Madison, WI 53711, United States}
\maketitle
\begin{abstract}
Mass spectrometry is a widespread approach to work out what are the constituents of a material. Atoms and molecules are removed from the material and collected, and subsequently, a critical step is to infer their correct identities based from patterns formed in their mass-to-charge ratios (m/z) and relative isotopic abundances. However, this identification step still mainly relies on individual user's expertise, making its standardization challenging, and hindering efficient data processing. Here, we introduce an approach that leverages modern machine learning technique to identify peak patterns in time-of-flight mass spectra within microseconds,
outperforming human users without loss of accuracy. Our approach is cross-validated on mass spectra generated from different time-of-flight mass spectrometry (ToF-MS) techniques, offering the ToF-MS community a open-source, intelligent mass spectra analysis. 

\end{abstract}

\section{Introduction}
Mass spectrometry is a widespread approach for revealing what constitutes a solution or a material. An array of techniques are used in the life sciences, in geology and materials science. Amongst this arsenal, time-of-flight mass spectrometry (ToF-MS) is one of the mainstream technique, in which an ion's mass-to-charge ratio is determined via a time-of-flight measurement \cite{Wolff1953}. It can provide quantitative analysis of the composition of the sampled material with a high precision and for a wide range of  atomic and molecular masses \cite{Maher2015}. The principles of ToF-MS are common to techniques such as matrix-assisted laser desorption/ionization (MALDI), secondary ion mass spectrometry (SIMS) or atom probe tomography (APT). Each of these techniques relies on a different concept to emit the ions from the sample and this versatility means that their common underlying analysis approach viz. ToF-MS, has found use in the studies of chemical reaction, large molecule characterization but also the quantification of dopants in semiconductors or the atomic scale distribution of impurities at grain boundaries in metallic alloys for instance \cite{Sulzer2012,Pedersen1359,Tanaka1988,Kissel1987,Karas1988,Liebscher2018}. 


The ToF-MS data is essentially a plot of the counts as a function of the mass-to-charge ratio – typically a peak appears for each isotope of each of the element present – and the amplitude is proportional to the relative amount of each species within the sampled volume. Fast and accurate identification and interpretation of the rich patterns and correlations in the spectral data are of great importance and can lead to new discoveries \cite{Aebersold2016}. Yet the interpretation and identification relies on the user's expertise, making it slow, prone to error and hindering reproducibility.

\begin{figure}[H]
\centering
	\includegraphics[width=1\textwidth]{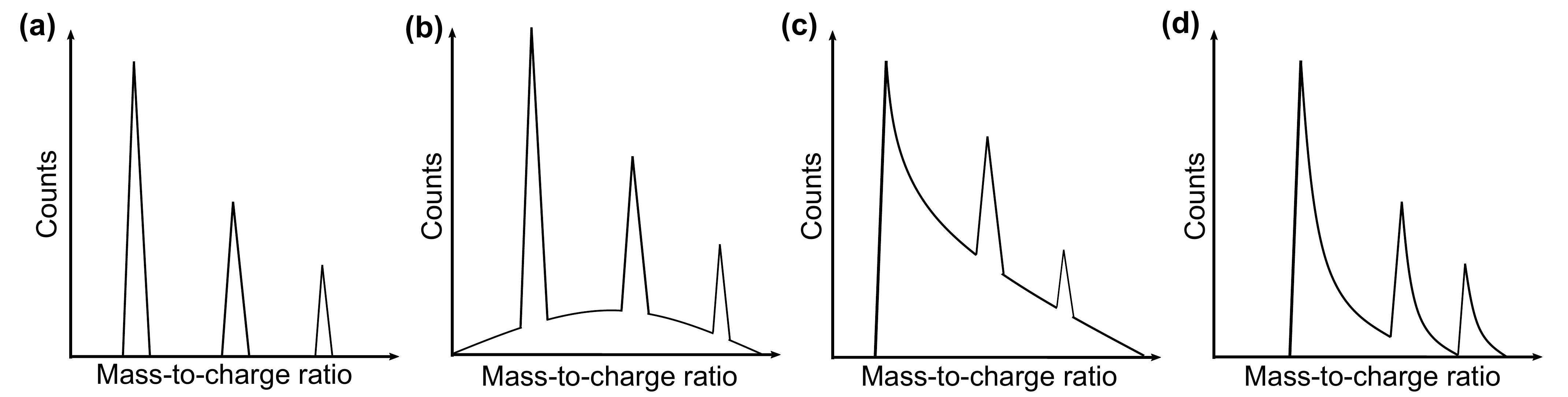}
	\caption{Examples of peak patterns under various experimental conditions \textbf{(a)} Perfect peak pattern. \textbf{(b)} peak pattern with background, due to experimental noise source. \textbf{(c)} Peak patterns with long thermal tails. \textbf{(c)} Peak patterns with short thermal tails.}
	\label{Peak_shape}
\end{figure}

Challenges in the development of automatic ToF-MS data analysis are two-fold. First, in ToF-MS, ions of the same species typically show a distribution in their velocity or a distribution in their instant of departure from the specimen. These lead to a distribution in flight times. As a result, depending on different experimental conditions, ToF-MS peak patterns can take various shapes and are not always simple to recognize (Figure \ref{Peak_shape})\cite{Ulrich2017}. Second, molecular patterns are commonly encountered in ToF-MS spectra, i.e. not only signals from atomic ions are detected \cite{Tsong1984f,Sha1992,Muller2011,Gordon2012,Rusitzka2018}. 
Combining individual atoms into a molecular ion usually leads to a new pattern comprising the distribution of the combination of isotopes from each individual element. Building a database for all possible molecular formula is practically impossible.

Machine learning is well-known for its powerful ability to recognize patterns and signals\cite{Jordan2015}. Recently, the mass spectrometry community has embraced machine learning techniques for large-scale data analysis. The data analysing speed of ion-trap-based mass spectrometry has been dramatically accelerated \cite{Elias2004, Gessulat2019}, whereas ToF-MS data analysis still largely relies on database searching \cite{Sadygov2004, Sinitcyn2018}. 

Some pioneering works demonstrated the potential of applying statistical/machine learning techniques to ToF-MS spectra analysis. For example, unsupervised ML has been used in the exploratory data analysis for ToF-SIMS and ToF-MALDI \cite{Biesinger2002, McCombie2005, Bluestein2016,Verbeeck2020}.

Lately a Bayesian approach has been adopted for peak identification in APT \cite{vurpillot_019, MIKHALYCHEV2020}. The Bayesian approach implemented by A. Mikhalychev et al. \cite{MIKHALYCHEV2020}  is able to identify and deconvolute many different types of ToF-APT mass spectra simultaneously. However, it still requires prior information from users and strong assumptions on the shape of the mass peaks.  

\begin{figure}[H]
\centering
	\includegraphics[width=1\textwidth]{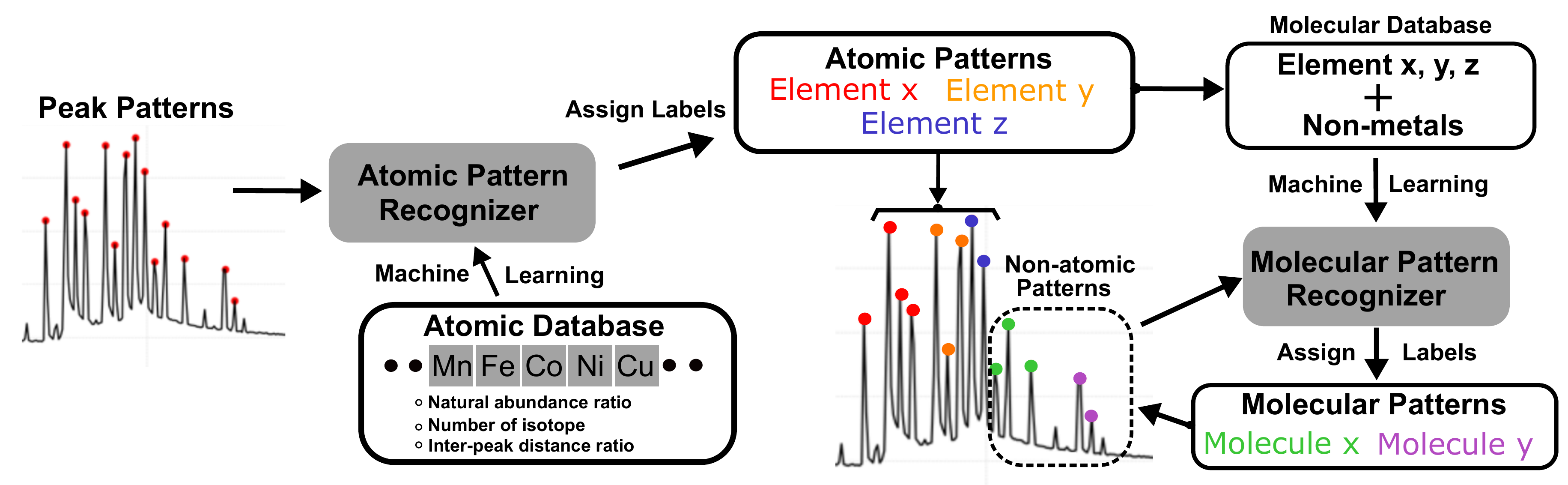}
	\caption{Flowchart of ML-assisted Time-Of-Flight mass spectrum identification (ML-ToF): Atomic pattern recognizer takes a mass spectrum as input and identifies all atomic patterns (mainly pure metal elements). Then a molecular database is constructed by combining atomic patterns from elements with non-metal elements (e.g. hydrogen, oxygen, nitrogen). Trained on such on-the-fly database, a machine-learning-based molecular pattern recognizer assigns molecular identities to non-atomic patterns. In such way, ML-ToF recognizes both the elemental and molecular fingerprints in mass spectra. }
	\label{Flowchart}
\end{figure}

Here, we introduce a machine-learning-based approach that automates the process of assigning elemental and molecular identities to peaks and series of peaks within ToF-MS spectra. Moreover, uncertainties are attached to these identities indicates to which extent the peak patterns are affected by the noise level and shape features. We name this approach 'ML-ToF'. It is shown that ML-ToF can handle various TOF-MS spectra, without a prior knowledge of composition information and from the analysis of a variety of materials systems and techniques. Indeed, we cross-validate ML-ToF on ToF-APT and ToF-SIMS spectra. The materials investigated include a high-strength Al-alloy developed for aerospace applications, medium-Mn steel found in automotive applications, Cu-In-based materials used in solar cell absorbers, and SmCo-based permanent magnets. Furthermore, we benchmark the results by comparing ML-ToF-assigned labels with those yielded by field experts. ML-ToF drastically reduces the duration of the peak recognition process. In general, it takes ML-ToF microseconds to obtain a labelled spectrum, whereas users could take minutes or even hours. An overview of our approach is shown in Figure \ref{Flowchart}.

\section{Peak Pattern Detection}
\begin{figure}[H]
\centering
	\includegraphics[width=0.8\textwidth]{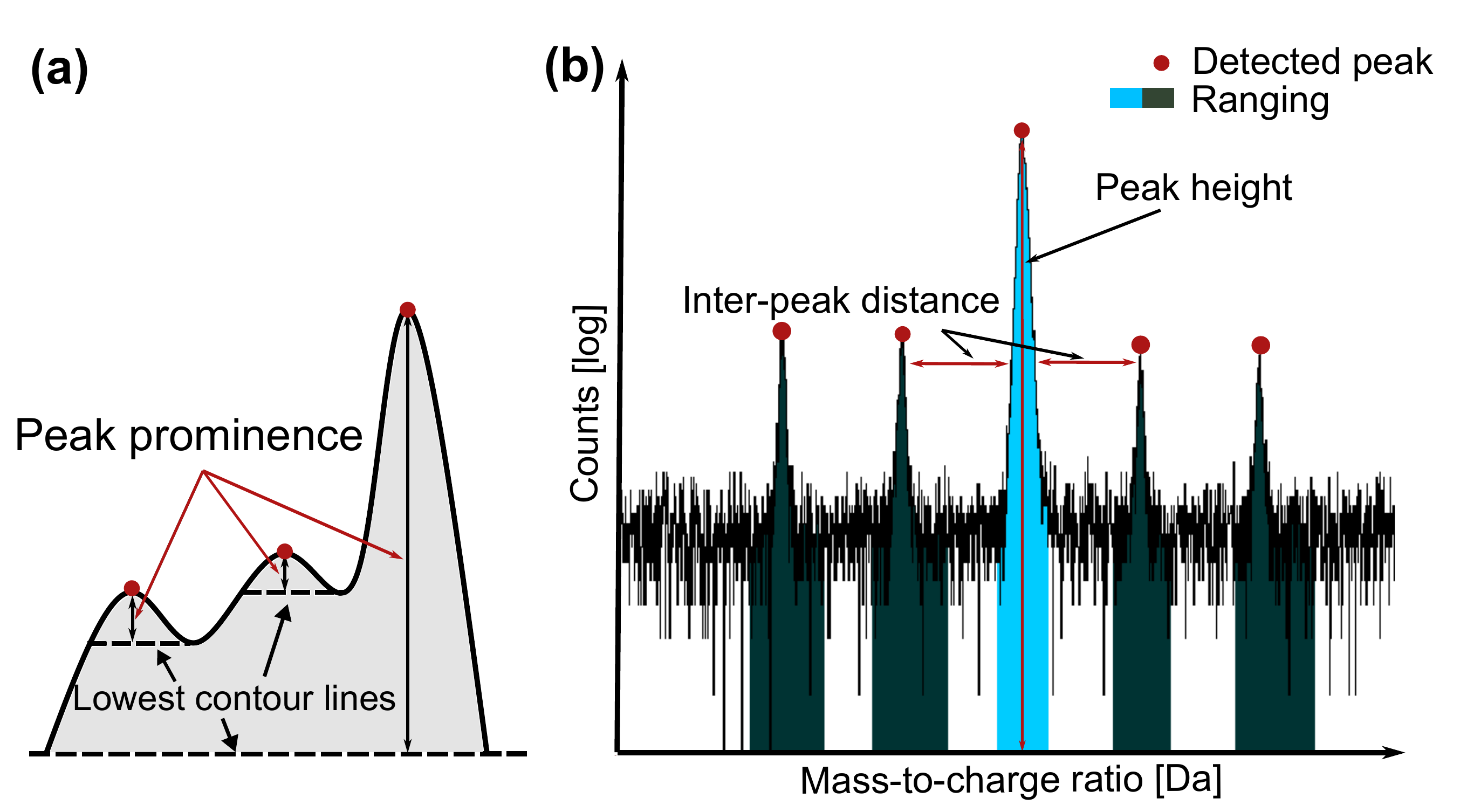}
	\caption{Examples of peak detection parameters in the Scipy python package \textbf{(a)} Schematic diagram showing the defintion of peak prominence: Peak prominence is defined as the vertical distance between the peak and the lowest contour line(the dashed lines). \textbf{(b)} Peak detection example from the ToF-APT dataset,  showing the inter-peak distance between detected peaks and peak height. Blue and dark green regions represent the the range of peaks assigned by human users.}
	\label{Peak_detection}
\end{figure}

Mass spectra can be regarded as a one-dimensional array whose values are always positive. We focus here on patterns with sufficient signal-to-background level to demonstrate that our approach can work properly with detectable patterns. We import the peak detection algorithm from a Python library (Scipy package, de facto standard package for signal processing in python) that finds the peak positions and the corresponding intensity values \cite{2020SciPy-NMeth}. The peak detection algorithm takes the mass spectra as input and searches for local maxima by simple comparison of intensity. A subset of these peaks can be further chosen by specifying conditions of peak properties. There are three major peak properties: peak height, inter-peak distance, and peak prominence. The prominence is defined as the intensity difference between the peak's height and its adjacent local minima, as indicated by Figure \ref{Peak_detection}(a). In Figure \ref{Peak_detection}(b) one can find the definition of peak height (the absolute count value in log scale). Throughout ToF-APT examples we used the same parameters for the detection (see Figure \ref{Peak_detection}): Peak height = 4 [log count]; Inter-peak distance = 0.25 Da; Prominence = 0.5 [log count]. By visual inspection, the peak detection algorithm with this set of parameters can capture the vast majority of peaks. 

In the manual procedure, users need to select a start and end position for each peak, as shown in Figure \ref{Peak_detection}(b). This procedure is often referred to as 'ranging' \cite{Hudson2011}, and this process can lead to errors due to the different of peak shapes, which depend in part on the instrument used, but also on the experimental conditions. For instance, the laser pulse energy or the base temperature were shown to have an influence \cite{Yao2011a,Tang2010a,LaFontaine2015a}. Here, we confine the task of ML-ToF to the identification of elemental or molecular patterns and assume the peak intensity is represented by the intensity at the detected position instead of the entire peak range. In practice, this assumption works well: ML-ToF can recognize the peaks even when they exhibit long tails. Tails originate either from energy deficits or uncertainty on the instant at which the ion left the specimen's surface \cite{Muller1974, Vurpillot2006b,Vurpillot2009a,Gault2012n} (See Section \ref{Result}). The detected mass-to-charge ratios and the corresponding intensity serve as the input of ML-ToF. 

\section{ToF-MS Pattern Recognition}
In general terms, patterns existing in the mass spectra can be categorized in two types: 
\begin{itemize}
    \item atomic pattern exhibiting the natural abundance ratio of one particular element;
    \item  molecular pattern, formed by two or more elements with mixed abundance ratio distribution. 
\end{itemize}

In this section, we introduce a systematic approach that identifies both types simultaneously. Two main aspects are addressed, i.e. the strategy to construct a reasonable database, and the search and identification of the most probable patterns.





\subsection{Atomic Pattern Recognizer}
First, we introduce the atomic pattern recognizer designed to identify all the atomic patterns. 
The general protocol is demonstrated in Figure \ref{protocal}.
\subsubsection{Database}
Machine learning can produce optimal results only if it is trained on a good database. In our case, the atomic pattern database consists of three parts: number of isotope peaks, natural abundance ratio and inter-peak distance ratio. The inter-peak distance ratio is defined as the distance between two neighbouring peaks divided by the smallest neighbouring distance within a group of peaks. For example, Fe$^+$ has four peaks at 54, 56, 57, 58 Da. So the distance ratio is (56-54)/(58-57):(57-56)/(58-57):(58-57)/(58-57)= 2:1:1. As such, even if Fe is in the form of charge state 2 with four peaks at 27, 28, 28.5, 29 Da, the inter-peak distance ratio is still 2:1:1.  We do not have to impose any constraints on the specific charge state of the elements. This is important as the charge-state-ratio can vary significantly (i.e. element Fe can have 1+, 2+ or 3+ charge state) based on the experimental parameters and even within a single dataset \cite{Kingham1982}. 

The database contains the most commonly encountered elements with atomic number up to 40  (excluding the inert gases), plus some lanthanides. In total, it contains 47 elements and compounds, such as S$_2$ and C$_2$. These compounds are included, because some of the elements have a strong tendency to form molecular ions, as frequently observed experimentally. Further information regarding the database can be found in the supplemental information \ref{data}. 
\subsubsection{Inter-peak Distance Ratio filter}
\begin{figure}[H]
\centering
	\includegraphics[width=1\textwidth]{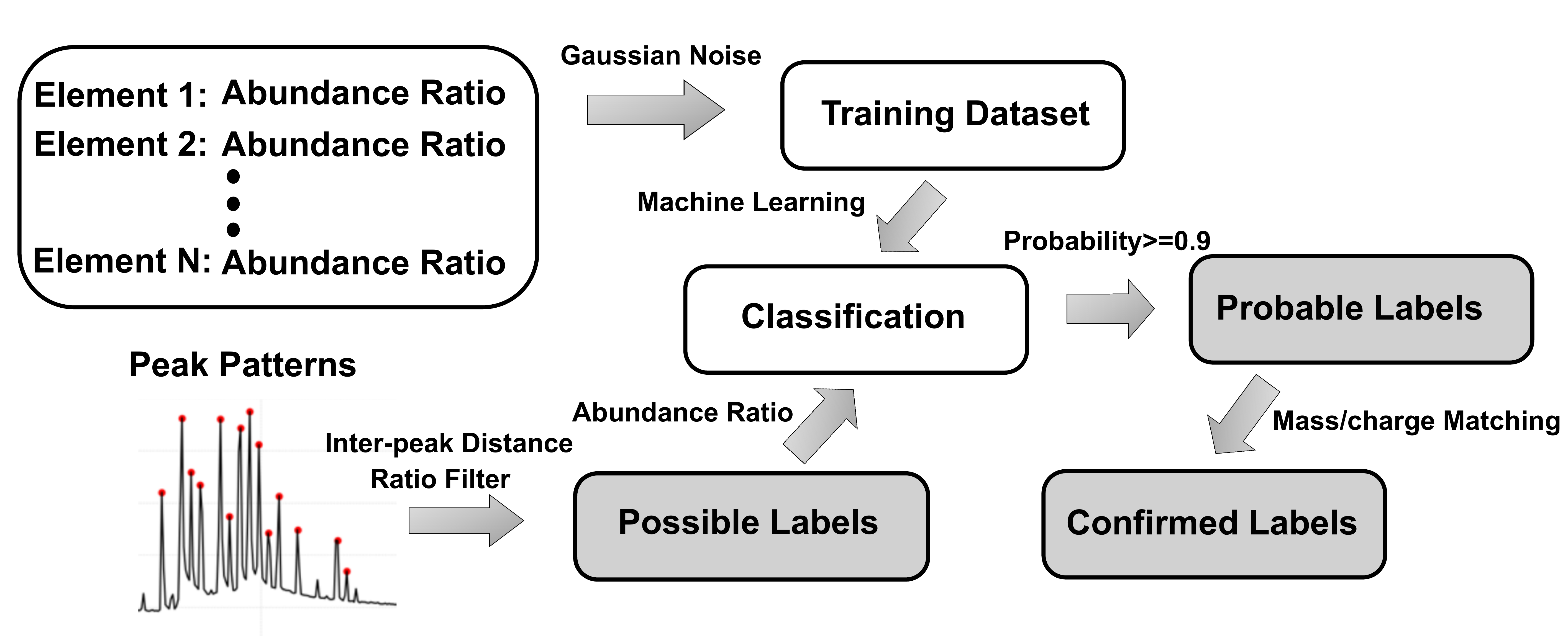}
	\caption{Protocol of atomic pattern recognizer. Patterns to be recognized are peaks with \textit{inter-peak distance ratio} and their respective abundance ratio. After the probable labels are obtained, a database search based on mass-to-charge is performed to identify the exact composition.}
	\label{protocal}
\end{figure}

As can be seen in Figure \ref{protocal}, matching the inter-peak distance ratio (IDR) is the first step towards a full pattern recognition. For a given peak pattern, the IDR filter searches for all possible candidates with matched IDR. Subsequently the algorithm will examine the abundance ratio of these candidates.

\subsubsection{Learning the Abundance Ratio}
\label{NAR}
The next step is concerned with pattern recognition of the isotopic abundance ratio. Classification of the abundance ratio is not a trivial task. Different patterns sometimes aggregate at a similar mass-to-charge ratio, it is often very difficult to deconvolute them. 
The ML technique is naturally suited for data-driven classification tasks, thanks to its ability to automatically learn and improve from experience without human intervention \cite{Jordan2015}. Unlike the conventional yes/no answer, ML algorithms produce a list of possible answers with corresponding likelihoods. In such cases even if an exact match from the given input to the theoretical database cannot be found, the ML-based algorithm can still provide a rank of likely labels. In other words, ML looks for partially retained patterns and thus assign a higher matching probability.

For elements with two isotopes, ML-ToF calculate the measured intensity ratio between the peaks ($r_m = P_1/P_2$) and compares to the expected ones from the natural abundances ($r_t$). 
If the absolute value of the deviation (r$_m$-r$_t$)/(r$_t$) exceeds certain threshold (here we chose empirically 0.3) then we classified this as unidentified peaks. For example the  pattern for Cu has a natural abundance ratio of 69.17:30.83, therefore the theoretical ratio $r_t$ = 69.17/30.83 = 2.24. ML-ToF will not assign element Cu to this pattern if its abundance ratio goes outside the range [1.56, 2.91].
For monoisotopic elements (e.g. Al, As, Co), since there is no abundance ratio,  ML-ToF searches for their different charge states and assigns the element if two or more of its corresponding charge states are found (e.g. Al$^+$ at 27 Da and Al$^{2+}$ at 13.5 Da). 

In the present study, we selected Light Gradient Boosting Machine (LightGBM) as our learning model. LightGBM belongs to the framework of Gradient Boosting Decision Tree (GBDT) \cite{Friedman2001}. GBDT is an ensemble model of weaker learners which are trained in sequence. In each training iteration, a decision tree learns from the errors up to the current iteration. Via a gradient descent approach, every subsequent tree minimizes the loss function between the actual output and the weighted sum of predictions from previous iterations. The final model is the weighted average of all weak learners. GBDT has achieved state-of-the-art performance in many machine learning tasks, such as multi-class classification \cite{Li2012} and ranking tasks \cite{Burges2010}.

Our label-predicting task is essentially a multi-label classification task. In such a setting, the algorithm tries to minimize the objective function $L$:
\begin{equation}
    L =  - \frac{1}{N}\left( \sum_{i=1}^{N} y_i\cdot \log (s_i) \right)
\end{equation}
$L$ represents the cross-entropy. Here, this ML specific entropy formulation serves as a measure for the difference between two probability distributions and is used as a loss functions for classification models; $N$ represents the number of labels; $y_i$ is the ground-truth and $s_i$ denotes predictions of ML model. This objective function measures how off the machines prediction is from the truth. The smaller the loss of objective function is, the closer the prediction of the machine is to the ground-truth. Zero loss would imply that the model has achieved 100$\%$ accuracy. In general, using the cross-entropy function instead of the sum of mean square errors for a classification problem leads to a faster training as well as improved generalization \cite{bishop2006}.
In contrast to other black-box ML models like a neural network, the decision tree enjoys a unique advantage, namely, it is an explainable ML model, which not only provides the predictions but also methods to interpret them. A specific example can be found in the supplemental information \ref{Model_Inter}.
Other parameters of the current LightGBM model and the corresponding explanations can be found in the supplemental information \ref{hyperparameters}.

\begin{figure}[H]
\centering
	\includegraphics[width=0.8\textwidth]{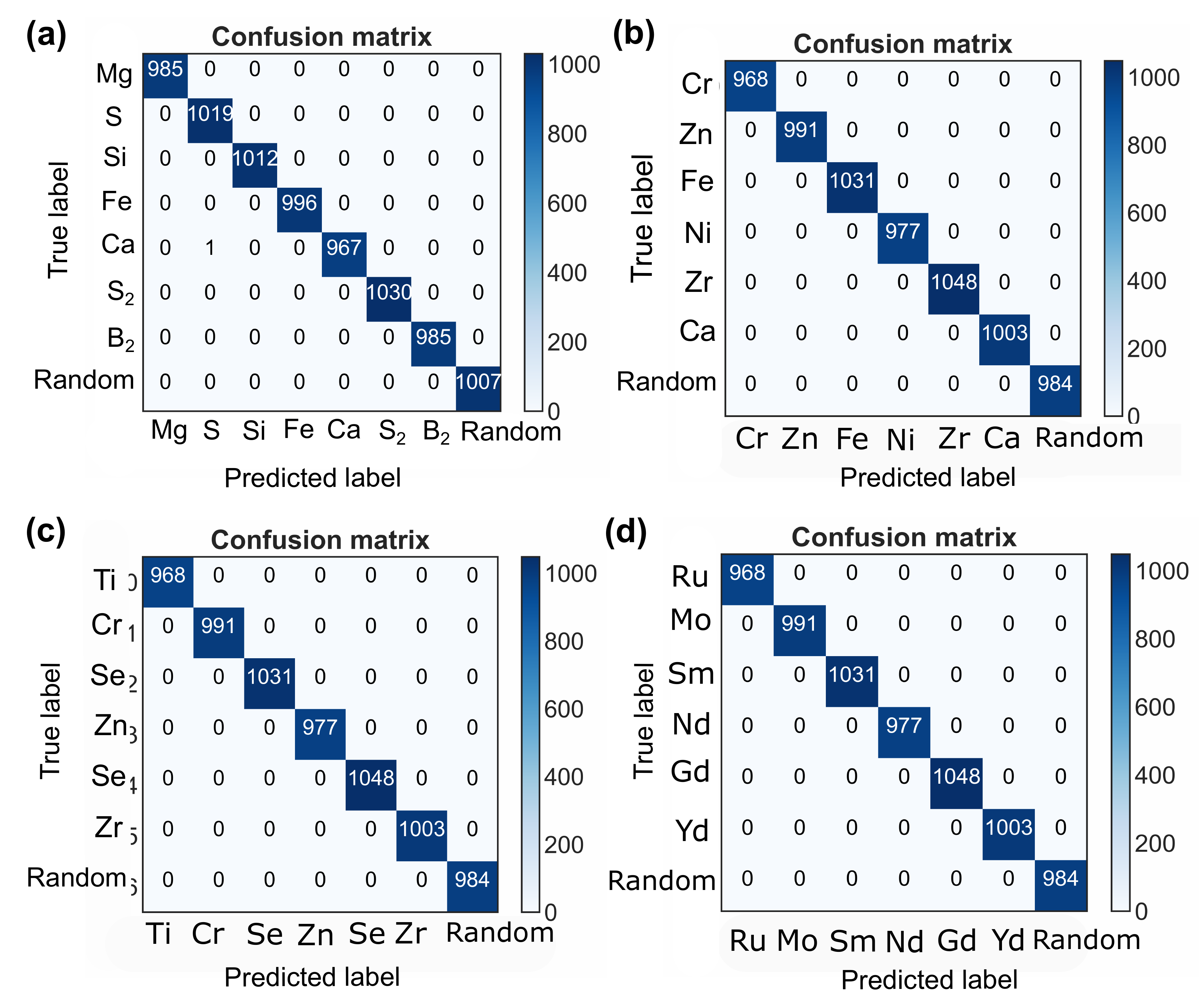}
	\caption{\textbf{(a, b, c, d)}  Confusion matrix for three, four, five and seven peak patterns: Confusion matrix indicates that models achieve 100$\%$ accuracy on abundance ratio classification task. A small randomness is introduced in the training/testing spliting. Therefore the size of test dataset is not always 1000 but quite close to it. }
	\label{confusion_matrix}
\end{figure}

We generate 5000 data points for each element. During the training the total dataset is further split into a first one used for the training (around 4000 data point) and the second (around 1000 data points) to validate the trained model. More detail of database construction can be found in the supplemental information \ref{data}.  Figure \ref{history}(a), (b), (c), (d) illustrate the training histories of the LightGBM model for three-, four-, five- and seven- peak patterns. The model for three-peak classification achieves near-zero loss after about 200 iterations, and then plateaus at zero. Loss histories of four-peak, five-peak and seven-peak patterns show similar trends. It is notable that the model for the four-peak pattern converges to zero at a much faster rate, reaching near-zero loss at 100 iteration. Thus this model stops early at 500 iterations. 
In all four cases, training and validating losses are almost identical, resulting in two completely overlapping curves.
\begin{figure}[H]
\centering
	\includegraphics[width=0.8\textwidth]{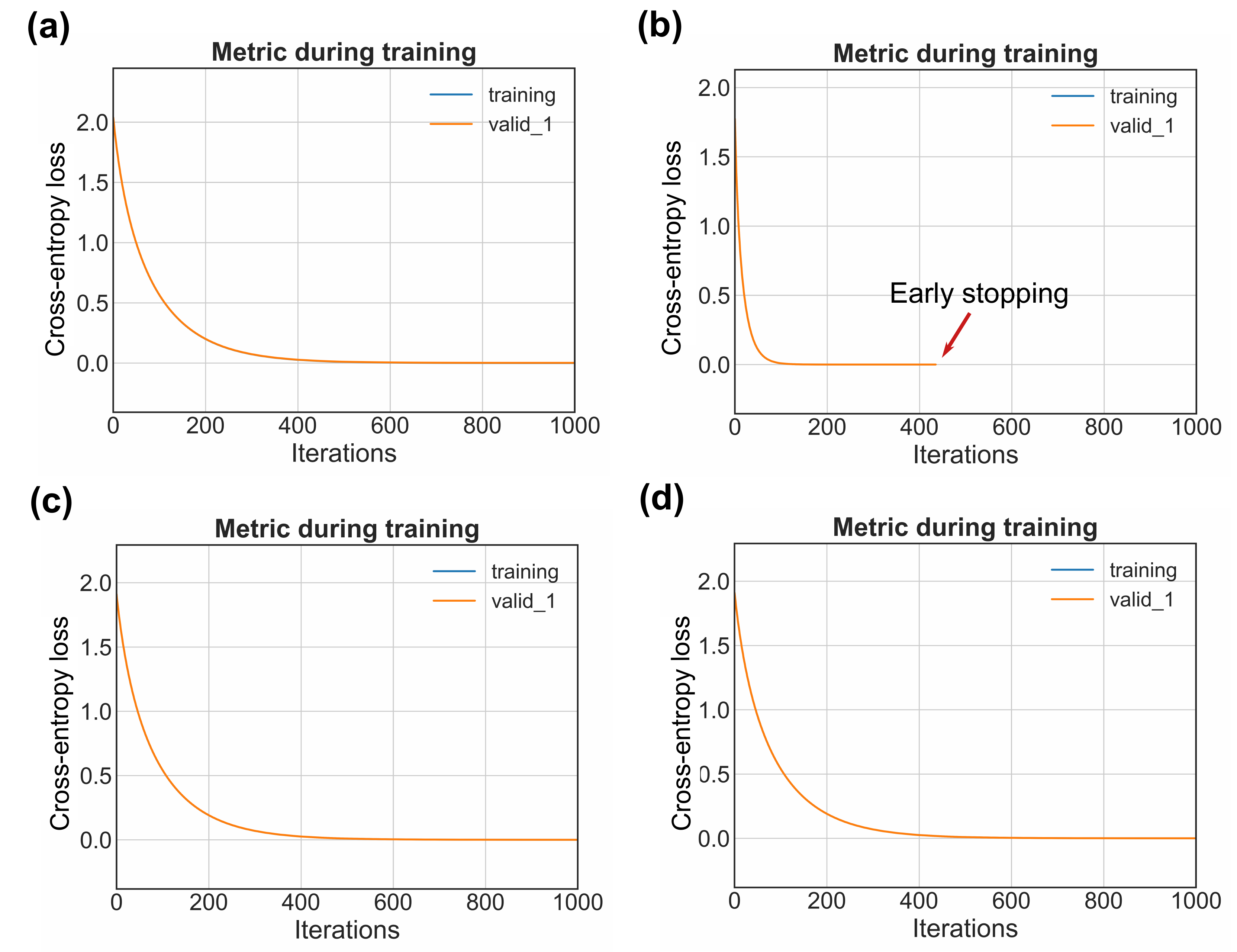}
	\caption{\textbf{(a, b, c, d)} Training histories of LightGBM model for three, four, five and seven peak pattern: Training histories of objective function $L$, we have training and validation curves (indicated by training and valid$\_$1 correspondingly). In all three cases, training and validating loss histories are almost the same. Hence the two curves overlap completely. }
	\label{history}
\end{figure}
The confusion matrix is a useful tool for visualizing the performance of a model. It enables a direct comparison between the ML prediction and ground-truth on test dataset.  These confusion matrices (shown in Figure \ref{confusion_matrix}(a), (b), (c), (d)) indicate that the LightGBM models can perfectly predict the element given its abundance ratio. In addition, the training dataset introduced 'redundancy' to deal with the partial pattern or overlapped pattern. For instance, three patterns are assigned to Fe: 1) Atomic mass: 54, 56, 57, 58 Da, abundance ratio: 5.8:91.8:2.1:0.3; 2) Atomic mass: 54, 56, 57 Da, abundance ratio: 5.8:91.8:2.1 and 3) Atomic mass: 56, 57, 58 Da, abundance ratio: 91.8:2.1:0.3. because sometimes the signal-to-noise ratio of some peaks is too weak to be detected. Or strong Ni presence (major peaks at 58 Da) destroy the first pattern of Fe, in these cases, ML-ToF is still be able to recognize the presence of Fe.


\subsubsection{Matching the Mass-to-charge Ratio}
A  'probable label' is defined as a  peak pattern with more than 90$\%$ certainty (assigned by LightGBM model). However, the probable label is not yet the final identified label. 
For example, if a pattern satisfies both IDR and abundance ratio of element Fe, it is still possible that this pattern can be other element. Therefore, as the last step, the probable label is confirmed only if its mass-to-charge ratio can be matched to a mass-to-charge ratio database, i.e. a pattern which has the same IDR and abundance ratio of an element. In the case of Fe, for instance, this would be if its m/z ratio is 54 Da, 56 Da, 57 Da, 58 Da, then ML-ToF predicts Fe$^+$, but if if its m/z ratio is 60 Da, 72 Da 73 Da, 74 Da, ML-ToF predicts FeO$^+$. 
\subsection{Molecular Pattern Recognizer}
\begin{figure}[H]
\centering
	\includegraphics[width=1\textwidth]{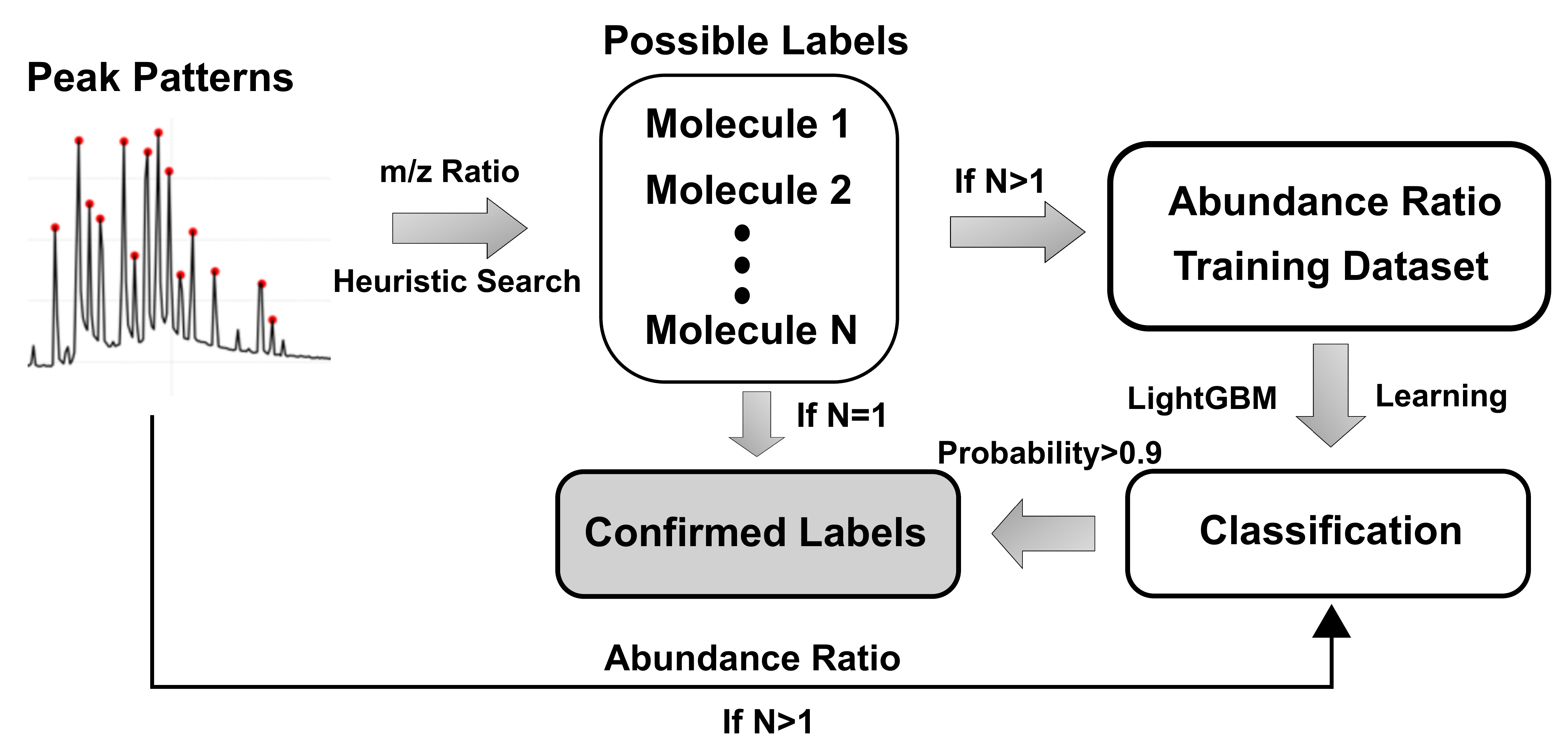}
	\caption{Molecular pattern recognizer}
	\label{MPR}
\end{figure}

When two or more elements with a different natural abundance ratio combine with each other, the resulting molecule forms a new fingerprint. As we mentioned in the introduction,  the new fingerprint differs not only in atomic number but also in the abundance ratio. This type of combination is often found between nonmetal element (e.g. carbon, oxygen, nitrogen, sulfur) and sometimes in metallic elements too \cite{Tsong1986}. This poses a significant challenge to the construction of the database, since it is impossible to search for all combinations by brute force. In order to identify the molecular fingerprint, we introduce molecular pattern recognizer, which adopts a different workflow comparing to the atomic pattern recognizer, as outlined in Figure \ref{MPR}.

For any undetermined patterns, molecular pattern recognizer firstly perform a heuristic search (Figure \ref{MPR}) by matching their mass-to-charge ratios to a on-the-fly molecular label database, and assign molecular label to this pattern if a match is found. This on-the-fly database contains all possible recombinations between the identified atomic patterns and non-metal elements. Range of this new molecular database is depending on the maximum detected mass-to-charge ratio. If there are multiple possible candidates, an abundance-ratio-based LightBGM will be trained and find out the most probable labels. This part is similar to the atomic pattern recognizer.

\section{Results and Discussion}
\label{Result}
\subsection{Atom Probe Tomography}
Atom probe tomography is a microscopy and microanalysis technique that provides three-dimensional compositional mapping of materials at the near-atomic scale \cite{MILLER2012158,Larson2013b,Muller2011}. Accurate analysis of atom probe data typically involves assigning an elemental nature to each ion based on its mass-to-charge-ratio in the ToF-APT mass spectrum. In this section, we evaluate the performance of our  approach on ToF-APT spectra from different alloy systems.

\subsubsection{Aerospace high-strength Al-alloy}
\begin{figure}[H]
\centering
	\includegraphics[width=1\textwidth]{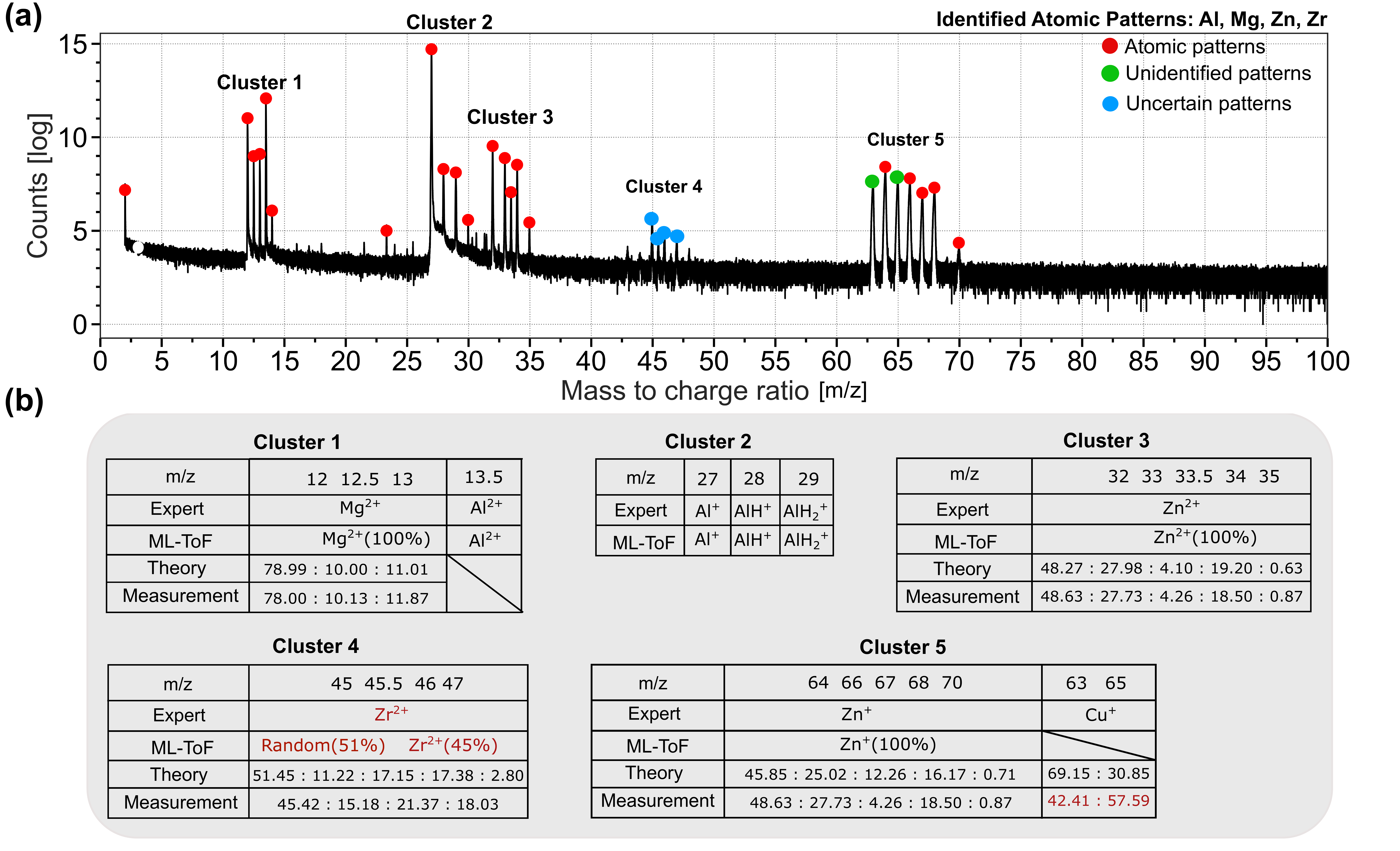}
	\caption{ML-ToF Identification of simple alloy system. \textbf{(a)} Ion mass spectrum of a simple alloy system: Color of circle marker indicates the state of peaks. Red, green blue markers indicate atomic(identified), unidentified and uncertain peaks, respectively; Majority of the ML-ToF assigned labels are consistent with APT operators. \textbf{(b)} Peak identity analysis: Five columns can be found for each individual cluster: mass-to-charge ratio; Expert-assigned element, ML-ToF-assigned element; Theoretical normalized intensity (Theory); Measured normalized intensity (Measurement).}
	\label{example_1}
\end{figure}

Al-Zn-Mg-Cu-(Zr) alloys are widely employed in aerospace and automobile applications due to their low mass density and high strength ability \cite{STARKE1996,Mondolfo2013}. These alloys are strengthened by a high volume fraction of nano-scale precipitates \cite{Dumont2005, Zhao2018}. ToF-APT of this alloy system generally has clear peak patterns and involves only a few molecular ions (demonstrated in Figure \ref{example_1}). In this first example, there are three possible categories for these detected peaks: Identified peaks, unidentified peaks and uncertain peaks. Overall, the patterns identified by ML-ToF are consistent with the expert's indexing and the ML-ToF-identified peaks account for 99.9$\%$ of the total intensity of detected patterns. 
To facilitate visualization, the peaks are grouped into five clusters and they are separately described in Figure \ref{example_1}(b). We provide a list of tables that compare expert-assigned elements to those assigned by ML-ToF. For cluster 1, 3, 4, 5 theoretical and measured normalized intensity (all involved normalized intensities sum up to 100) are also present.
More specifically, one can observe that for clusters 1, 3, 5, ML-ToF and expert are in complete agreement, ML-ToF assigns 100$\%$ certainty to its selected candidates (shown in the bracket after the assigned element). Whereas in cluster 4 (mass-to-charge ratio: 45 Da, 45.5 Da, 46 Da, 47 Da), the ML algorithm is confused between a random (51$\%$) and Zr pattern(45$\%$). There are two main reasons leading to this result. The first relates to the detection criteria: the intensity of the fifth peak is too low such that peak at 48 Da is not detected. The second relates to the abundance ratio: the measured abundance ratio significantly differs from the natural abundance ratio of Zr. The normalized intensity of second peak (in theory the percentile is 11.22$\%$ but measured to be 15.18$\%$) deviates 36$\%$ from theory. This deviation is likely originated from the detection of Zr-H peaks \cite{Mouton2019}. Despite the uncertainty, ML-ToF still ranks Zr as the second most likely candidate with 45$\%$ certainty.  

Moreover, in the case of the green colored peaks within cluster 5, ML-ToF is not able to assign any identity to peak patterns with mass-to-charge ratio values of 63 and 65 Da, whilst the expert would assign them as Cu$^+$. This is owing to the fact that ML-ToF  makes predictions of two peak patterns based on a simple threshold method. In this case the measured intensity ratio between the two peaks is 0.73. Meanwhile if it is standing-alone element Cu, this ratio would be 2.24. Hence ML-ToF observe a remarkable deviation (67.1$\%$) and rejects candidate Cu, contrary to the assignment of expert. CU in its 1+ charge state is also prone to being detected as a CuH$_2^{1+}$, which will then lead to CuH$_2$ to overlap with the Zn peak at 67 Da, which in parts, explains the discrepancy between the measured and theoretical ratios for Zn, which did not affect ML-ToF's capacity to identify Zn correctly. 

\subsubsection{Medium-Mn steel}
\begin{figure}[H]
\centering
	\includegraphics[width=1\textwidth]{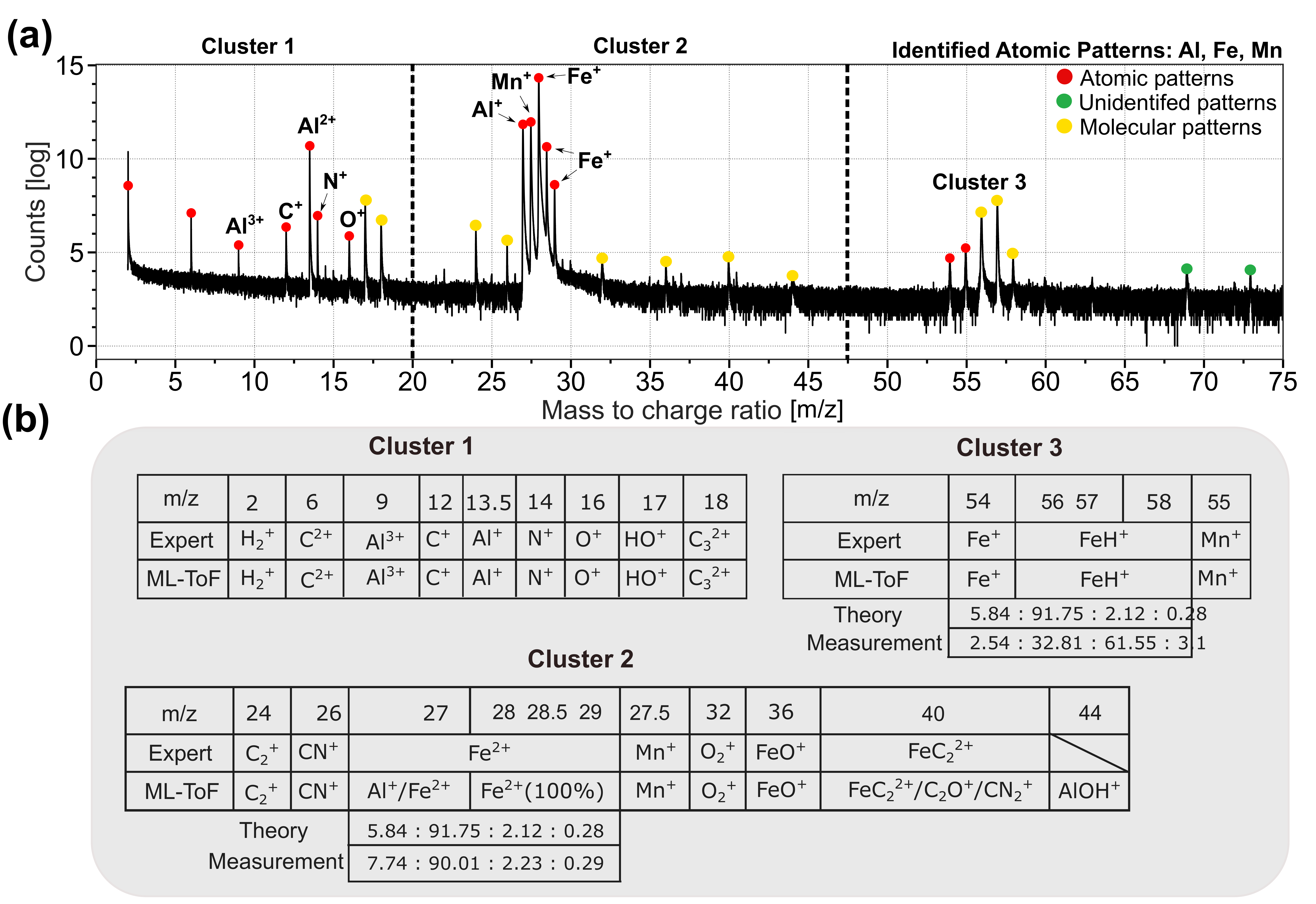}
	\caption{Identification of Fe-Mn-C-Al alloy system. \textbf{(a)} Ion mass spectrum of Fe-Mn-C-Al alloy: markers are colored based on the indicates the state of peaks (red for identified and green unidentified, yellow suggest molecular ions); dashed lines are used to separate clusters; peaks identified by atomic pattern recognizer are indicated. Majority of the peaks are identified by ML-ToF. Among which atomic patterns constitute 98$\%$ intensity of detected peaks and about 1$\%$ are of possible Molecule origins. \textbf{(b)} Peak identity analysis: The patterns by which expert and ML-ToF do not agree each other are emphasized using red text.}
	\label{example_2}
\end{figure}

Medium-manganese steels are promising candidates for the automotive industry owing to their excellent mechanical properties \cite{lee_han_2015}. Atom probe studies help understand the local chemistry, in particular the crystal defects such as dislocations and grain boundaries \cite{kuzmina_herbig_ponge_sandlobes_raabe_2015, kuzmina_ponge_raabe_2015, KwiatkowskidaSilva2018, KWIATKOWSKIDASILVA2019109}, thereby providing insights into the atomic scale mechanisms at play in these class of steels. Figure \ref{example_2} (a) illustrates a mass spectrum for the more complex Fe-Mn-C-Al alloy system.
More than 99$\%$ of the ions are within detected peaks that were assigned an identity that is consistent with that given by field expert. 

ML-ToF successfully identified the existence of element Fe, Mn, Al. Nonmetal elemental patterns of O, N, C are identified too. Therefore, a new database is proposed, which contains four different types of molecular patterns: Fe$_x$H$_a$C$_b$N$_c$O$_d$, Al$_x$H$_a$C$_b$N$_c$O$_d$, Mn$_x$H$_a$C$_b$N$_c$O$_d$ and H$_a$C$_b$N$_c$O$_d$. The number of metal (x) is set to 1, 2, 3, 4; H(a) to 0, 1, 2; C(b), N(c), O(d) to 0, 1, 2, 3, 4 and  charge state to 1, 2.  These ranges includes almost all the common types of molecular patterns. Additionally, the search of molecular patterns is restricted to values below 70 Da, since no peaks occur beyond this value. Combing all the above-mentioned conditions, we construct a molecular pattern database shown in Table 2. 

\begin{table}[h!]
\centering
\caption{Molecular pattern database: x = 1, 2 ; a = 0, 1, 2; b = 0, 1, 2, 3, 4; c = 0, 1, 2, 3, 4; d = 0, 1, 2, 3, 4; charge state = 1, 2 and mass-to-charge ratio is restricted to below 75 Da, since no peaks are detected beyond such. The search of molecular pattern is performed within this dataset. }
\begin{tabular}{ P{2.5cm}||P{2.5cm}|P{2.5cm}|P{2.5cm}|P{2.5cm} }
 \hline
 Molecular pattern  & Fe$_x$H$_a$C$_b$N$_c$O$_d$ & Al$_x$H$_a$C$_b$N$_c$O$_d$ & Mn$_x$H$_a$C$_b$N$_c$O$_d$  & H$_a$C$_b$N$_c$O$_d$\\
 \hline

 Database size & 329 & 455 & 222 & 750\\
 \hline
\end{tabular}
\label{databse_1}
\end{table}


Figure \ref{example_2}(b) shows both the expert's and ML-ToF's assignment of peaks. In cluster 1, ML-ToF fails to identify peak Al$^{3+}$ at 9 Da due to a limitation of the current database - i.e. for the moment no 3+ charge state are taken into account.
In cluster 2, both Al$^+$ and Fe$^+$ were assigned to the peak at 27 Da, which is a known overlap that makes the quantification by APT of Al in Fe or Fe in Al challenging. Even in the presence of Al, atomic pattern recognizer is still able to recognize the Fe isotope pattern with 100$\%$ certainty. At 40 Da, the algorithm  offers some multiple candidates (FeC$_2^{2+}$ CN$_2^+$,C$_2$O$^+$, with the same number of atoms) as compared to the expert's choice of FeC$_2^{2+}$. In such a case, the algorithm would also choose FeC$_2^{2+}$ since Fe is the most abundant element (80$\%$ of intensity are assigned to element Fe).

\subsubsection{Sm-Co-based hard magnet}
Sm-Co-based materials are known for their outstanding magnetic properties, which are related to their complex microstructure \cite{Maury1993,Gutfleisch2011}. By changing the pinning mechanisms and pinning strength, the coercivity of the alloy \ce{Sm2(Co, Fe, Cu, Zr)_17} can be controlled by substituting \ce{Fe} for \ce{Co} \cite{Duerrschnabel}. In this example,  ML-ToF shows its robustness against broadened peaks due to the relatively high laser power used for this analysis.
\begin{figure}[H]
\centering
	\includegraphics[width=1\textwidth]{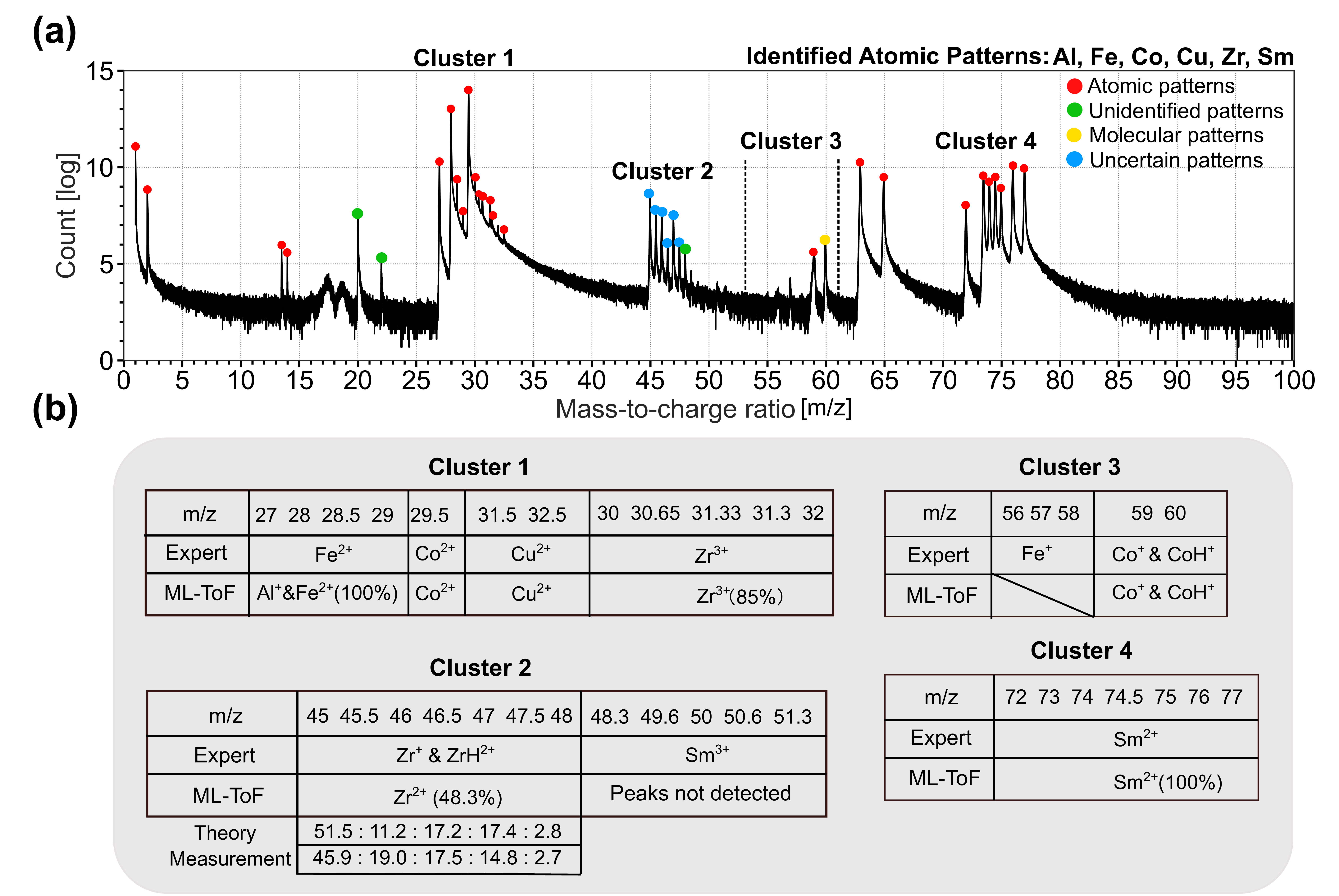}
	\caption{Identification of Sm-Co-based hard magnet. \textbf{(a)} APT Mass spectra  of  a Sm-Co-based hard magnet; \textbf{(b)} Table of identity analysis for detected peaks.}
	\label{example_4}
\end{figure}
In cluster 1, ML-ToF  identified aluminium due to detection of peaks at \ce{Al^+} (peak at 13.5 Da) and Al$^{2+}$ (peak at 27 Da).  Also, ML-ToF identifies Zr$^{3+}$, albeit with reduced certainty (85$\%$), this is likely due to the long thermal tails of the peaks.
In cluster 2, ML-ToF identified Zr$^{2+}$ with 48.3$\%$ certainty at 45 Da, 45.5 Da, 46 Da, 47 Da, 48 Da. This relatively low probability (still considerably higher than the second highest pattern: random (30$\%$)) indicates the existence of other type of ions, which is pointed out by expert as ZrH$^2+$. ML-ToF fails to assign any labels to peaks at 48.3 Da, 49.6 Da, 50 Da, 50.6 Da, 51.3 Da. This is largely due to their relatively low signal-to-background ratio and thus do not meet our detection criteria. In the cluster 3, peaks at 56 Da, 57 Da, 58 Da are not detected due to their low signal-to-noise ratio but still labeled by experts as Fe$^+$. Finally at cluster 72 Da, 73 Da, 74 Da, 74.5 Da, 75 Da, 76 Da, 77 Da, the element Sm is identified.

In fact, elemental signatures like N$^+$(peak at 14 Da), As$^+$ (peak at 75 Da), Sc$^+$ (peak at 45 Da), Ca$^{2+}$ are identified too. But since we did not detect other charge states from these one/two peak elements, ML-ToF rejects these possible candidates. This can be considered as an inherent limit of the instrument itself rather than ML-ToF. 

\subsubsection{Solar cell absorber}
Here, we showcase ML-ToF's application to a much more complex mass spectrum. Cu(In,Ga)S$_2$ (CIGS) is a compound semiconductor with a direct band gap, which can be tuned between 1.55 eV to 2.4 eV for pure CuInS$_2$ and CuGaS$_2$, respectively \cite{Scheer2011}. It is therefore suitable as an absorber material in solar cells, especially as a top junction in tandem solar cells to overcome the Shockley-Queisser limit \cite{DeVos1980}. However, the microstructure, especially the composition-structure relationships of grain boundaries, for this material is not well known \cite{Lomu2019, SCHWARZ2020105081}. Here, we present for clarity only the mass spectrum of the Cu-In-S system (without Ga).
\begin{figure}[H]
\centering
	\includegraphics[width=1\textwidth]{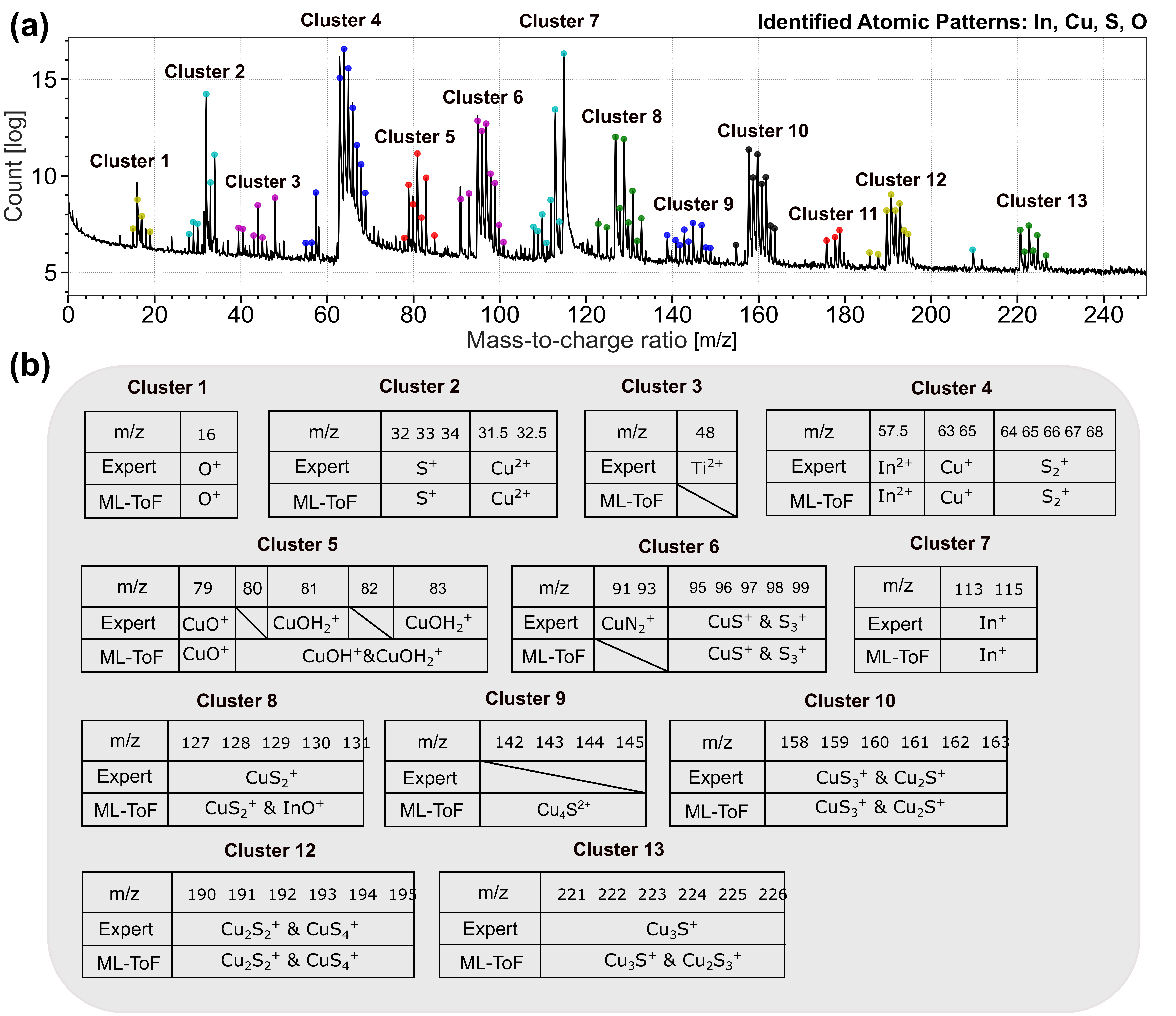}
	\caption{Identification of Cu-In-S system. \textbf{(a)} Ion mass spectrum of a Solar cell absorber system. Since most of the peaks are molecular pattern, for the better visualization, circular markers with different colors are used to separate different clusters. Atomic pattern recognizer has identified In, Cu, S, O as the atomic elements. \textbf{(b)} Peak identity analysis.}
	\label{example_3}
\end{figure}

Indexing the complex mass spectrum, shown in Figure \ref{example_3}(a), is more difficult than the previous two cases. ML-ToF identifies atomic fingerprints: In, Cu, S and O. As they tend to recombine with each other, the newly formed molecular pattern will not only change in terms of the atomic number but also their abundance ratio. Such an example is shown in Table 2(a). Cu and S forms a compound (CuS) with atomic number of 95, 97, 99 and new abundance ratio of 63.7 : 32.2 : 1.3. Nevertheless, as we can see in the peak identity analysis in Figure \ref{example_3}(b). Without any prior knowledge, ML-ToF provides almost identical result as the field expert.

\begin{figure}[H]
\centering
    	\caption*{Table 2: \textbf{(a)} An example of new molecule pattern formation. Molecule CuS shows a new patterns.\textbf{(b)} New database: x = 1, 2,3, 4 ; y = 1, 2, 3; a= 0,1,2,3; b = 0, 1, 2; charge state = 1, 2 and mass-to-charge ratio is restricted to below 300 Da, since no peaks are detected beyond that. The search of molecular patterns is performed within this dataset.}
	\includegraphics[width=0.7\textwidth]{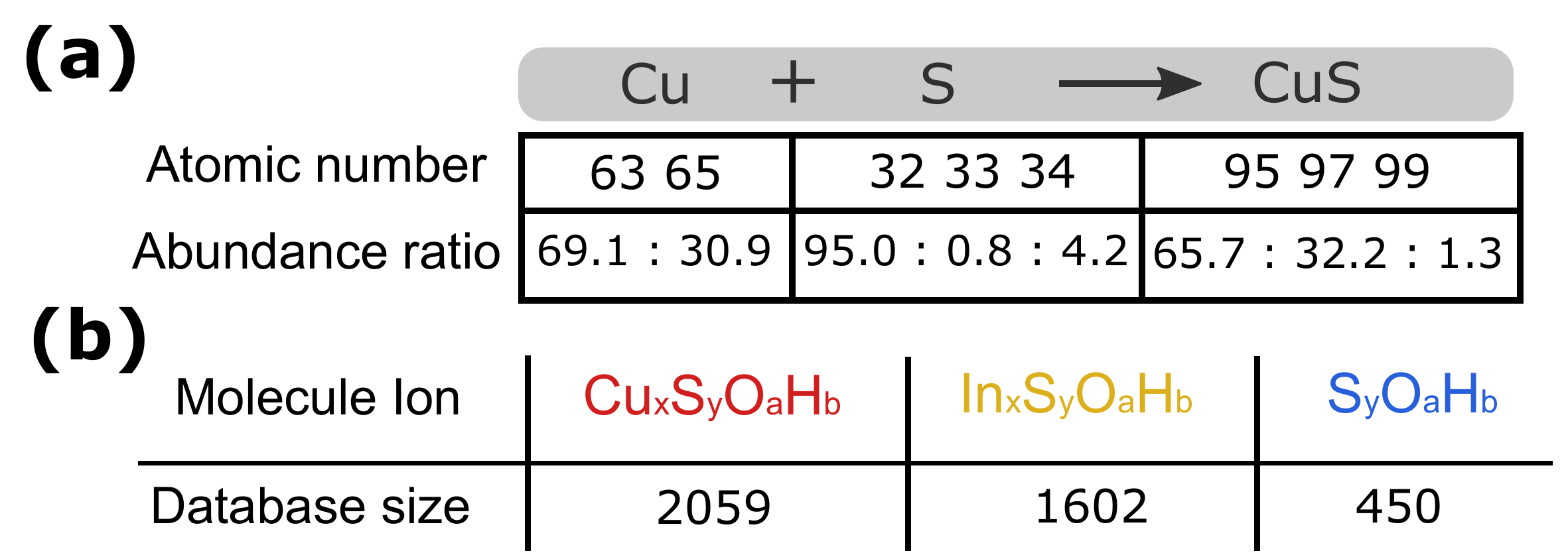}
	\label{database_2}
\end{figure} 

For cluster 1 - 4, cluster 8, cluster 10, cluster 12 and cluster 14, ML-ToF's choice of element identity is identical. For cluster 3, ML-ToF fails to assign any labels to peak 48 Da whereas the expert assign Ti$^+$.
This is owing to the fact that the background signal is relatively higher comparing to the side peaks of Ti, therefore only one peak is detected whereas in theory element Ti should show five peaks. Regarding cluster 5 (81 - 83 Da), the expert chose CuOH$_2^+$ while ML-ToF chose CuOH$^+$ and CuOH$_2^+$.

There are two other interesting cases that are worth mentioning. The first case is CuN$_2^+$, which is identified at (91-93 Da, cluster 6) but not confirmed by ML-ToF. A closer look reveals that this ambiguity is due to the fact that ML-ToF did not identify the pattern associated to nitrogen at 7 Da or 14 Da, i.e. N$^2{+}$ and N$^+$. Therefore no N-containing compounds in the new molecular pattern database involves nitrogen. In the second case, ML-ToF is able to predict the identity (Cu$_4$S$_2^+$ and InS$^+$) at 142 - 145 Da (cluster 9) whereas the user did not assign any identity to them. 

Overall, ML-ToF has shown high fidelity handling the complicated cases, even identifying some peaks by which human did not assign any label. More importantly, it takes ML-ToF only half second to complete the task, which by experts would have taken 15 minutes on average - somtimes even longer when scientists had no prior experience with the material system.

\subsection{Secondary Ion Mass Spectrometry}
ToF-SIMS is another analytical imaging mass spectrometry technique, which provide unique insights into surface
chemistry \cite{Liebl1967,WITTMAACK197539,Magee1978}. The large-scale and high-dimensional data generated by contemporary ToF-SIMS instruments consists of x-y-z spatial information and mass spectrum associated with each pixel. The strength of SIMS, in comparison to APT, is the sensitivity associated to the larger probe volumes. The associated drawback is the lower spatial resolution. A single ToF-SIMS dataset contains hundreds to thousands mass spectra. In comparison to APT mass spectra, peak patterns of ToF-SIMS generally has high signal-to-noise ratio. 
In spite of the fact that, many peak patterns have very low intensity, these peaks are still of great importance and thus need to be identified. Hence the detection criteria is also different from ToF-APT: Peak height = 0.0001 [log count]; inter-peak distance = 0.25 Da; prominence = 0.0001 [log count]. In the following examples, we demonstrate the efficiency of ML-ToF on on ToF-SIMS mass spectra of different complexity. Here we omit the tabular peak analysis and directly insert ML-ToF assigned-labels. As the expert-assigned labels are only available for a few peaks.

\subsubsection{Corrosion and wear Co-based alloy}
\begin{figure}[H]
\centering
	\includegraphics[width=1\textwidth]{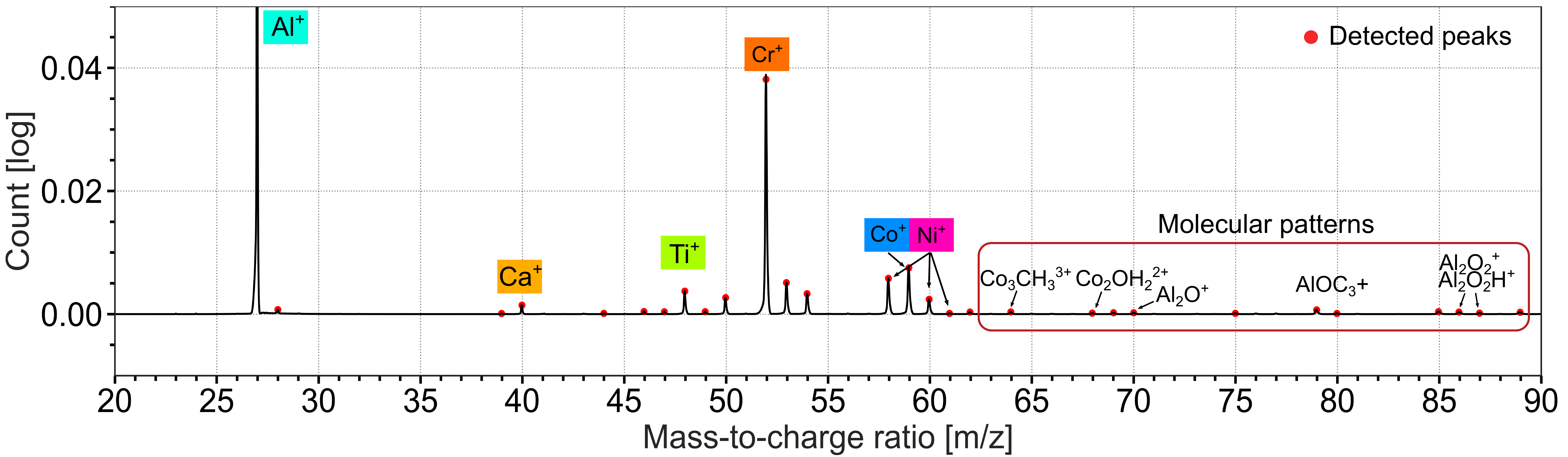}
	\caption{Identification of spectral patterns from Secondary ion mass spectrometry using ML-ToF.  Region of interest of Mass-to-charge ratio ranges from  20 Da to 90 Da. ML-ToF also identifies complex molecular patterns from 60 to 90 Da. }
	\label{example_SIMS_2}
\end{figure}

The chemical composition (wt.$\%$) of this alloy characterized by Nanoscopic Secondary Ion Mass Spectrometry (Nano-SIMS) is Ni-0.32, Cr-0.20, Al-0.08, Y-0.4, balanced with Co, which is designed as a corrosion and wear resistant alloy employed in turbine blades \cite{Yang2020}. 
The mass spectrum shown in Figure \ref{example_SIMS_2} was constructed by TOF-SIMS Explorer 1.3.1.0 software from the total ions information of the scanned region. In this spectrum, ToF-ML identifies  Al$^+$, Cr$^+$,  Co$+$, Ni$+$, Ca$^+$, Ti$^+$. This composition is relatively simple. However, abundant complex molecular fingerprints are identified by ML-ToF, as evidenced in Figure \ref{example_SIMS_2}. 

\subsubsection{Unknown alloy from mine dump}
Finally, ML-ToF was tested on an unknown alloy sample from a mine dump in Erzgebirge, Germany. there is no specification for nominal composition. The spectrum is produced by dynamic SIMS, showing complex peak patterns. ToF-ML identifies a variety of elements and compounds: Na$^+$, Al$^+$, Fe$^+$, Co$^+$, Cu$^+$, Ni$^+$, As$^+$, Mo$^+$, Bi$^+$, NaO$^+$, MnO$^+$, CuO$^+$. ML-ToF is able to extract rich information even with no prior knowledge on the material.
\begin{figure}[H]
\centering
	\includegraphics[width=1\textwidth]{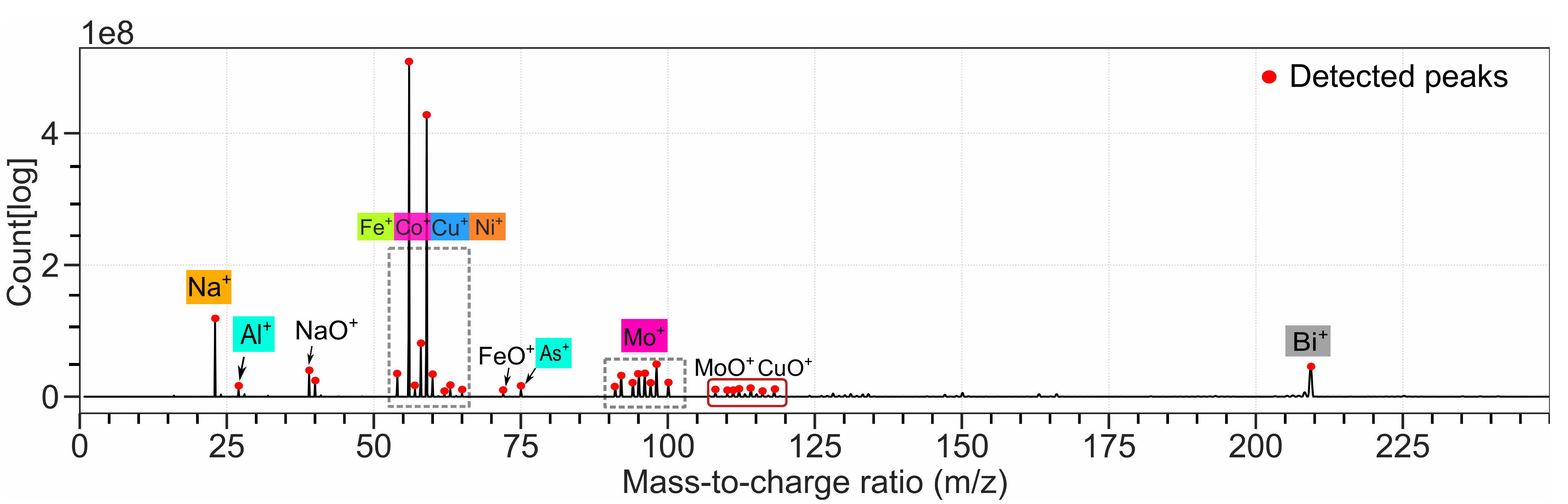}
	\caption{ML-ToF successfully assigns labels to vast majority of peaks from mass spectra of an unknown alloy sample from a mine dump in Erzgebirge, Germany.}
	\label{example_SIMS_1}
\end{figure}



\section{Conclusions}
We have developed a gradient-boosting-decision-tree-based approach which converts raw time-of-flight mass spectra to its elemental of molecular identified form. The training dataset is generated based on natural abundance ratios which does not require any human labeling. The workflow is validated on experimental datasets from APT and SIMS. Its outputs are compared with identification provided by different operators. 
The main bottleneck of our approach mainly lies at the detection limits. Higher signal-to-noise ratio of the spectrum will lead to more identified patterns. The training dataset does not include all elements in the periodic table. Because sufficient testing and validation must be performed when new elements are added to the training data. Mass spectra containing these new elements were not typically available at the time the method was being developed. The next step is to collect more data and  extend ML-ToF to more element type, thus making ML-ToF a universal technique for ToF spectral data analysis. Currently, ML-ToF still relies on brute-force search of molecular ion combination, to accelerate this search process,  one could envision a heuristic search algorithm to be integrated into the ML-ToF (e.g. Beam search \cite{Reddy1977}), which rules out impossible combination of ions. Finally, the implementation of real-time ML-ToF for mass spectra patterns recognition during the atom probe experiment has the potential of avoiding peak overlapping problem, thus further boost the accuracy of APT.
Finally, our method is open-source, easy to implement and capable of making instant, accurate and consistent predictions. A wide range of ToF-based techniques can be benefited from this approach, e.g. hunting for patterns of biomarkers in high-throughput ToF-MALDI data or for contamination on the solid surface in SIMS data etc. MT-ToF enables significant acceleration of the identification process and paves the way for more reliable and more reproducible data analysis.





\section{Acknowledgements}
BG acknowledges the support from the SFB TR 270 - Hommage, Project Z01. YW appreciates the funding by BiGmax, the Max Planck Society's Research Network on Big-Data-Driven Materials Science. The authors are grateful to Profs. Leopoldo Molina-Luna and Oliver Gutfleisch  for the provision of the Sm-Co hard magnet sample.

\section{Supplementary Information}
To ensure transparency and reusability the entire program is written in Python and is available in Github: \url{https://github.com/DeepHeisenberg}.
\addtocontents{toc}{\protect\setcounter{tocdepth}{0}}
\subsection{Data Generation}
\label{data}
The four-peak pattern database is shown in Table 3 as an example. 
\begin{table}[H]
\centering
\caption*{Table 3: Cr as an example which is included in the database. All relevant information can be found in the standard element table. With 'random' class included in the database, machine can learn to classify those severely distorted patterns as 'random'. }
\begin{tabular}{ |P{2cm}||P{2.5cm}|P{3.5cm}|P{4cm}|  }
 \hline
 Elements & Number of Isotopes &Inter-peak Distance Ratio &  Abundance Ratio\\
 \hline
 Cr & 4 & 2:1:1 & 4.3 : 83.8 : 9.5 : 2.4  \\ \hline
\end{tabular}
\end{table}
The full database is also available in Github: \url{https://github.com/DeepHeisenberg}.

The real experimental datasets always contains noise. To handle this problem, we introduce noise to the training dataset. In the field of machine learning this technique often refers to  'data augmentation'. The data with noise is generated according to the formula: 
\begin{equation}
      X \sim \mathcal{N}(\mu,\,\sigma^{2})\,.
\end{equation}
X is the generated data. $\mathcal{N}$ is the normal distribution. $\mu$ is the natural abundance ratio and $\sigma$ is chosen to be $0.01 \cdot \mu$. 
We generate 5000 data point per element  and split it into training and  testing set at ratio of 80:20. 

Machine learning algorithm requires fixed size of input. Therefore we have separate classification tasks depending on the number of peaks. Based on the present database, our current LightBGM models constitute three-peak (e.g. Mg has isotopes at 24, 25, 26, so in total three peaks.) , four-peak, five-peak, seven-peak classification.

\subsection{LightGBM Model Hyperparameters}
\label{hyperparameters}
LightGBM uses the leaf-wise tree growth algorithm using gradient descent method \cite{LightBGM}.  The hyperparameters of LightGBM model used in ML-ToF are the following: 
\begin{itemize}
    \item Number of leaves: 50
    \item minimum data in leaf: 20
    \item Max depth: 10
    \item Learning rate: 0.005
    \item Bagging fraction: 0.9
    \item Feature fraction: 0.9
    \item Bagging frequency: 5
\end{itemize}
Number of leaves is one of the most important parameters that controls the complexity of the tree model. It sets the maximum number of leaves each weak learner. Max depth control the max depth of each trained tree. Bagging fraction is to specify the percentage (0.9) of rows used per tree building iteration. Feature fraction is designed to randomly select a subset (0.9) of features on each iteration (tree). Bagging frequency setting at 5 means the perform bagging at every 5 iteration.   These hyperparameters are chosen by a few rounds of trial-and-error.  A complete explaination of these hyperparameters can be found in the Ref. \cite{LightBGM}. 
\subsection{LightGBM Model Interpretation}
\label{Model_Inter}
\begin{figure}[H]
\centering
	\includegraphics[width=1\textwidth]{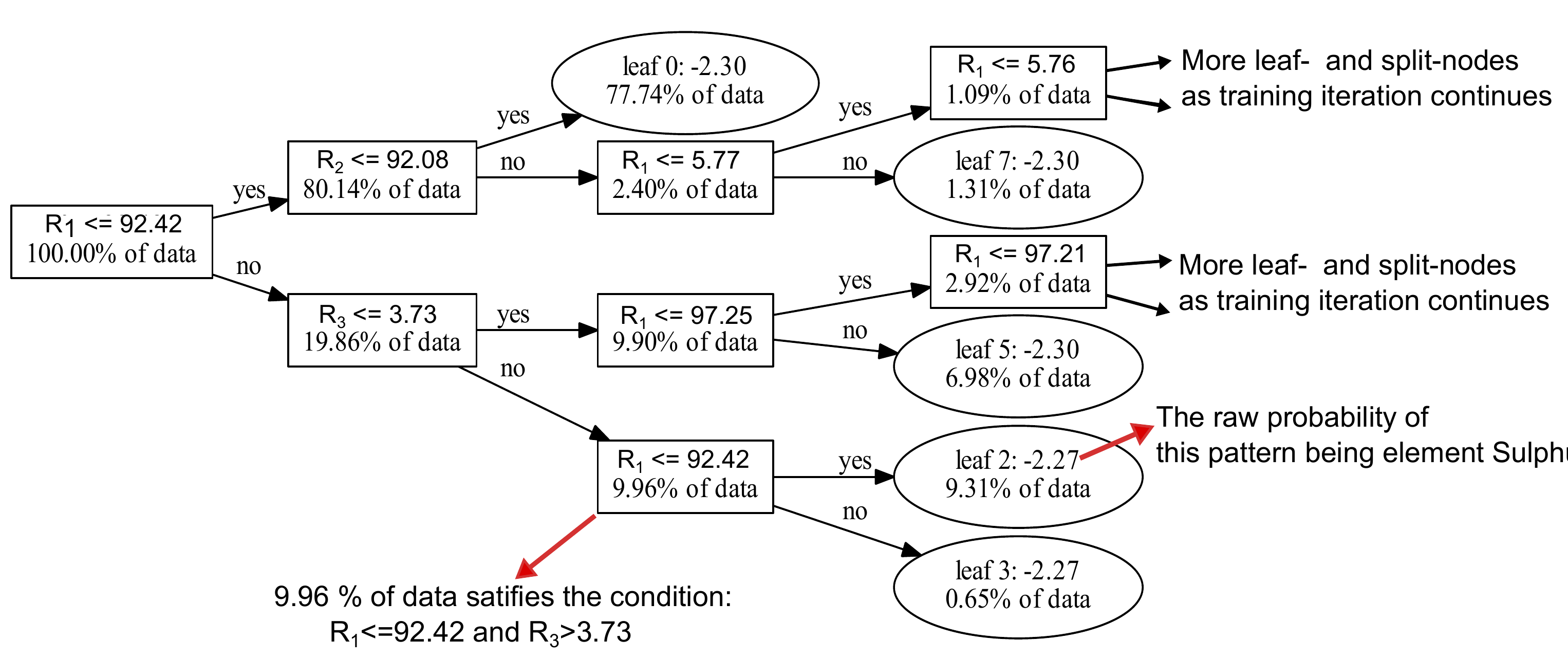}
	\caption{The first Fe classifier for three-peak pattern: P(1,2,3) denote the intensity of peaks in Equations; std stands for standard deviation; The ellipses indicate the leaf nodes, where the tree stops further splitting; The rectangles indicate the decision node, where a node splits into further sub-nodes. Each leaf node is featured by the raw probability of the peak pattern being element Fe (before the applying the equation \ref{eq:3}) and percentile of training data belongs to this node. }
	\label{Learn_DT}
\end{figure}

The first three-peak pattern classifier for element Sulphur is illustrated in Figure \ref{Learn_DT}, from which we can see how the tree model makes decision quantitatively. For instance, for the training data, 9.31$\%$ of them can be assigned to leaf 2, because the first and third peak ratio R$_1$ and R$_3$ are smaller than 92.42 and 3.37 respectively . Given these features, the probability of this pattern being element Sulphur is:
\begin{equation}
    P(Sulphur) =  \frac{1}{1+exp(-(-2.27))} = 0.0936
    \label{eq:3}
\end{equation}
The equation above is sigmoid function that converts the raw probability (-2.27) to the normalized probability (sum of probabilities equals to 1). The first classifier is a weak classifier, since it is at the very beginning of the training . The chance it identifies the right pattern closes to random(10$\%$). As the training iteration increases, the tree grows and ensemble prediction accuracy (weighted probability from all weak classifiers) on testing dataset reaches 100$\%$ . Results are shown as confusion matrix( in Figure \ref{confusion_matrix}(e)). 

\bibliographystyle{ieeetr}
\bibliography{main}

\begin{thebibliography}{10}

\bibitem{Wolff1953}
M.~M. Wolff and W.~E. Stephens, ``{A pulsed mass spectrometer with time
  dispersion},'' {\em Review of Scientific Instruments}, vol.~24, no.~8,
  pp.~616--617, 1953.

\bibitem{Maher2015}
S.~Maher, F.~Jjunju, and S.~Taylor, ``Colloquium : 100 years of mass
  spectrometry: Perspectives and future trends,'' {\em Reviews of Modern
  Physics}, vol.~87, pp.~113--135, 01 2015.

\bibitem{Sulzer2012}
P.~Sulzer, F.~Petersson, B.~Agarwal, K.~H. Becker, S.~J{\"{u}}rschik, T.~D.
  M{\"{a}}rk, D.~Perry, P.~Watts, and C.~A. Mayhew, ``{Proton transfer reaction
  mass spectrometry and the unambiguous real-time detection of 2,4,6
  trinitrotoluene},'' {\em Analytical Chemistry}, vol.~84, no.~9,
  pp.~4161--4166, 2012.

\bibitem{Pedersen1359}
S.~Pedersen, J.~L. Herek, and A.~H. Zewail, ``The validity of the "diradical"
  hypothesis: Direct femtoscond studies of the transition-state structures,''
  {\em Science}, vol.~266, no.~5189, pp.~1359--1364, 1994.

\bibitem{Tanaka1988}
K.~Tanaka, H.~Waki, Y.~Ido, S.~Akita, Y.~Yoshida, T.~Yoshida, and T.~Matsuo,
  ``{Protein and polymer analyses up to m/z 100 000 by laser ionization
  time‐of‐flight mass spectrometry},'' {\em Rapid Communications in Mass
  Spectrometry}, vol.~2, no.~8, pp.~151--153, 1988.

\bibitem{Kissel1987}
J.~Kissel and F.~R. Krueger, ``{The organic component in dust from comet Halley
  as measured by the PUMA mass spectrometer on board Vega 1},'' {\em Nature},
  vol.~326, no.~6115, pp.~755--760, 1987.

\bibitem{Karas1988}
M.~Karas and F.~Hillenkamp, ``Laser desorption ionization of proteins with
  molecular masses exceeding 10,000 daltons,'' {\em Analytical Chemistry},
  vol.~60, no.~20, pp.~2299--2301, 1988.
\newblock PMID: 3239801.

\bibitem{Liebscher2018}
C.~H. Liebscher, A.~Stoffers, M.~Alam, L.~Lymperakis,
  O.~Cojocaru-Mir{\'{e}}din, B.~Gault, J.~Neugebauer, G.~Dehm, C.~Scheu, and
  D.~Raabe, ``{Strain-Induced Asymmetric Line Segregation at Faceted Si Grain
  Boundaries},'' {\em Physical Review Letters}, vol.~121, no.~1, 2018.

\bibitem{Aebersold2016}
R.~Aebersold and M.~Mann, ``{Mass-spectrometric exploration of proteome
  structure and function},'' {\em Nature}, vol.~537, no.~7620, pp.~347--355,
  2016.

\bibitem{Ulrich2017}
U.~Boesl, ``{Time-of-flight mass spectrometry: Introduction to the basics},''
  {\em Mass Spectrometry Reviews}, vol.~36, pp.~86--109, jan 2017.

\bibitem{Tsong1984f}
T.~Tsong, ``{Pulsed-laser-stimulated field ion emission from metal and
  semiconductor surfaces: A time-of-flight study of the formation of atomic,
  molecular, and cluster ions},'' {\em Physical Review B}, vol.~30,
  pp.~4946--4961, nov 1984.

\bibitem{Sha1992}
W.~Sha, L.~Chang, G.~D. W. D.~W. Smith, E.~J.~J. Mittemeijer, C.~Liu, and
  E.~J.~J. Mittemeijer, ``{Some aspects of atom-probe analysis of Fe-C and Fe-N
  systems},'' {\em Surface Science}, vol.~266, pp.~416--423, apr 1992.

\bibitem{Muller2011}
M.~M{\"{u}}ller, D.~Saxey, G.~Smith, and B.~Gault, ``{Some aspects of the field
  evaporation behaviour of GaSb},'' {\em Ultramicroscopy}, vol.~111,
  pp.~487--492, may 2011.

\bibitem{Gordon2012}
L.~M. Gordon, L.~Tran, and D.~Joester, ``{Atom probe tomography of apatites and
  bone-type mineralized tissues},'' {\em ACS Nano}, vol.~6, no.~12,
  pp.~10667--10675, 2012.

\bibitem{Rusitzka2018}
K.~A.~K. Rusitzka, L.~T. Stephenson, A.~Szczepaniak, L.~Gremer, D.~Raabe,
  D.~Willbold, and B.~Gault, ``{An atomic-scale view at the composition of
  amyloid-beta fibrils by atom probe tomography},'' {\em Scientific Reports},
  vol.~8, no.~November, pp.~1--10, 2018.

\bibitem{Jordan2015}
M.~I. Jordan and T.~M. Mitchell, ``{Machine learning: Trends, perspectives, and
  prospects},'' {\em Science}, vol.~349, no.~6245, pp.~255--260, 2015.

\bibitem{Elias2004}
J.~E. Elias, F.~D. Gibbons, O.~D. King, F.~P. Roth, and S.~P. Gygi,
  ``{Intensity-based protein identification by machine learning from a library
  of tandem mass spectra},'' {\em Nature Biotechnology}, vol.~22, no.~2,
  pp.~214--219, 2004.

\bibitem{Gessulat2019}
S.~Gessulat, T.~Schmidt, D.~P. Zolg, P.~Samaras, K.~Schnatbaum, J.~Zerweck,
  T.~Knaute, J.~Rechenberger, B.~Delanghe, A.~Huhmer, U.~Reimer, H.~C. Ehrlich,
  S.~Aiche, B.~Kuster, and M.~Wilhelm, ``{Prosit: proteome-wide prediction of
  peptide tandem mass spectra by deep learning},'' {\em Nature Methods},
  vol.~16, no.~6, pp.~509--518, 2019.

\bibitem{Sadygov2004}
R.~G. Sadygov, D.~Cociorva, and J.~R. Yates, ``{Large-scale database searching
  using tandem mass spectra: Looking up the answer in the back of the book},''
  {\em Nature Methods}, vol.~1, no.~3, pp.~195--202, 2004.

\bibitem{Sinitcyn2018}
P.~Sinitcyn, J.~D. Rudolph, and J.~Cox, ``Computational methods for
  understanding mass spectrometry–based shotgun proteomics data,'' {\em
  Annual Review of Biomedical Data Science}, vol.~1, no.~1, pp.~207--234, 2018.

\bibitem{Biesinger2002}
M.~C. Biesinger, P.-y. Paepegaey, N.~S. McIntyre, R.~R. Harbottle, and N.~O.
  Petersen, ``Principal component analysis of tof-sims images of organic
  monolayers,'' {\em Analytical chemistry}, vol.~74, p.~5711—5716, November
  2002.

\bibitem{McCombie2005}
G.~McCombie, D.~Staab, M.~Stoeckli, and R.~Knochenmuss, ``Spatial and spectral
  correlations in maldi mass spectrometry images by clustering and multivariate
  analysis,'' {\em Analytical Chemistry}, vol.~77, no.~19, pp.~6118--6124,
  2005.
\newblock PMID: 16194068.

\bibitem{Bluestein2016}
B.~M. Bluestein, F.~Morrish, D.~J. Graham, J.~Guenthoer, D.~Hockenbery, P.~L.
  Porter, and L.~J. Gamble, ``{An unsupervised MVA method to compare specific
  regions in human breast tumor tissue samples using ToF-SIMS},'' {\em
  Analyst}, vol.~141, no.~6, pp.~1947--1957, 2016.

\bibitem{Verbeeck2020}
N.~Verbeeck, R.~M. Caprioli, and R.~{Van de Plas}, ``{Unsupervised machine
  learning for exploratory data analysis in imaging mass spectrometry},'' {\em
  Mass Spectrometry Reviews}, vol.~39, no.~3, pp.~245--291, 2020.

\bibitem{vurpillot_019}
F.~Vurpillot, C.~Hatzoglou, B.~Radiguet, G.~Da~Costa, F.~Delaroche, and
  F.~Danoix, ``Enhancing element identification by expectation–maximization
  method in atom probe tomography,'' {\em Microscopy and Microanalysis},
  vol.~25, no.~2, p.~367–377, 2019.

\bibitem{MIKHALYCHEV2020}
A.~Mikhalychev, S.~Vlasenko, T.~Payne, D.~Reinhard, and A.~Ulyanenkov,
  ``Bayesian approach to automatic mass-spectrum peak identification in atom
  probe tomography,'' {\em Ultramicroscopy}, vol.~215, p.~113014, 2020.

\bibitem{2020SciPy-NMeth}
P.~{Virtanen}, R.~{Gommers}, T.~E. {Oliphant}, M.~{Haberland}, T.~{Reddy},
  D.~{Cournapeau}, E.~{Burovski}, P.~{Peterson}, W.~{Weckesser}, J.~{Bright},
  S.~J. {van der Walt}, M.~{Brett}, J.~{Wilson}, K.~{Jarrod Millman},
  N.~{Mayorov}, A.~R.~J. {Nelson}, E.~{Jones}, R.~{Kern}, E.~{Larson},
  C.~{Carey}, {\.I}.~{Polat}, Y.~{Feng}, E.~W. {Moore}, J.~{Vand erPlas},
  D.~{Laxalde}, J.~{Perktold}, R.~{Cimrman}, I.~{Henriksen}, E.~A. {Quintero},
  C.~R. {Harris}, A.~M. {Archibald}, A.~H. {Ribeiro}, F.~{Pedregosa}, P.~{van
  Mulbregt}, and S.~.~. {Contributors}, ``{SciPy 1.0: Fundamental Algorithms
  for Scientific Computing in Python},'' {\em Nature Methods}, vol.~17,
  pp.~261--272, 2020.

\bibitem{Hudson2011}
D.~Hudson, G.~D. W. D.~W. Smith, and B.~Gault, ``{Optimisation of mass ranging
  for atom probe microanalysis and application to the corrosion processes in Zr
  alloys},'' {\em Ultramicroscopy}, vol.~111, pp.~480--486, may 2011.

\bibitem{Yao2011a}
L.~Yao, J.~M. Cairney, C.~Zhu, and S.~P. Ringer, ``{Optimisation of specimen
  temperature and pulse fraction in atom probe microscopy experiments on a
  microalloyed steel},'' {\em Ultramicroscopy}, vol.~111, no.~6, pp.~648--651,
  2011.

\bibitem{Tang2010a}
F.~Tang, B.~Gault, S.~P. S.~P. Ringer, and J.~M. J.~M. Cairney, ``{Optimization
  of pulsed laser atom probe (PLAP) for the analysis of nanocomposite Ti-Si-N
  films.},'' {\em Ultramicroscopy}, vol.~110, pp.~836--843, jun 2010.

\bibitem{LaFontaine2015a}
A.~{La Fontaine}, B.~Gault, A.~Breen, L.~Stephenson, A.~V. Ceguerra, L.~Yang,
  T.~{Dinh Nguyen}, J.~Zhang, D.~J. Young, and J.~M. Cairney, ``{Interpreting
  atom probe data from chromium oxide scales},'' {\em Ultramicroscopy},
  vol.~159, pp.~354--359, 2015.

\bibitem{Muller1974}
E.~W. M{\"{u}}ller and S.~V. Krishnaswamy, ``{Energy deficits in pulsed field
  evaporation and deficit compensated atom-probe designs},'' {\em Review of
  Scientific Instruments}, vol.~45, no.~9, pp.~1053--1059, 1974.

\bibitem{Vurpillot2006b}
F.~Vurpillot, B.~Gault, A.~Vella, M.~Bouet, and B.~Deconihout, ``{Estimation of
  the cooling times for a metallic tip under laser illumination},'' {\em
  Applied Physics Letters}, vol.~88, no.~9, p.~94105, 2006.

\bibitem{Vurpillot2009a}
F.~Vurpillot, J.~Houard, A.~Vella, and B.~Deconihout, ``{Thermal response of a
  field emitter subjected to ultra-fast laser illumination},'' {\em Journal of
  Physics D-Applied Physics}, vol.~42, no.~12, p.~125502, 2009.

\bibitem{Gault2012n}
B.~Gault, M.~P.~M. Moody, J.~M. Cairney, S.~S.~P. Ringer, J.~Cariney, and
  S.~S.~P. Ringer, {\em {Atom Probe Microscopy}}, vol.~160 of {\em Springer
  Series in Materials Science}.
\newblock New York, NY: Springer New York, springer s~ed., 2012.

\bibitem{Kingham1982}
D.~R. Kingham, ``{The post-ionization of field evaporated ions: A theoretical
  explanation of multiple charge states},'' {\em Surface Science}, vol.~116,
  pp.~273--301, apr 1982.

\bibitem{Friedman2001}
J.~H. Friedman, ``{Greedy function approximation: A gradient boosting
  machine},'' {\em Annals of Statistics}, vol.~29, no.~5, pp.~1189--1232, 2001.

\bibitem{Li2012}
P.~Li, ``Robust logitboost and adaptive base class {(ABC)} logitboost,'' {\em
  CoRR}, vol.~abs/1203.3491, 2012.

\bibitem{Burges2010}
C.~J.~C. Burges, ``{From rankNet to LambdaRank to lambdaMART: An overview},''
  {\em Learning}, vol.~11, pp.~23--581, 2010.

\bibitem{bishop2006}
C.~M. Bishop, {\em Pattern Recognition and Machine Learning (Information
  Science and Statistics)}.
\newblock Berlin, Heidelberg: Springer-Verlag, 2006.

\bibitem{Tsong1986}
T.~T. Tsong, ``{Observation of doubly charged diatomic cluster ions of a
  metal},'' {\em Journal of Chemical Physics}, vol.~85, no.~1, pp.~639--640,
  1986.

\bibitem{MILLER2012158}
M.~K. Miller, T.~F. Kelly, K.~Rajan, and S.~P. Ringer, ``The future of atom
  probe tomography,'' {\em Materials Today}, vol.~15, no.~4, pp.~158 -- 165,
  2012.

\bibitem{Larson2013b}
D.~J. Larson, T.~J. Prosa, R.~M. Ulfig, B.~P. Geiser, and T.~F. Kelly, ``{Local
  electrode atom probe tomography},'' {\em New York, US: Springer Science},
  p.~318, 2013.

\bibitem{STARKE1996}
E.~Starke and J.~Staley, ``Application of modern aluminum alloys to aircraft,''
  {\em Progress in Aerospace Sciences}, vol.~32, no.~2, pp.~131 -- 172, 1996.

\bibitem{Mondolfo2013}
L.~Mondolfo, {\em Aluminum alloys: structure and properties}.
\newblock Elsevier, 2013.

\bibitem{Dumont2005}
M.~Dumont, W.~Lefebvre, B.~Doisneau-Cottignies, and A.~Deschamps,
  ``{Characterisation of the composition and volume fraction of $\eta'$ and
  $\eta'$ precipitates in an Al–Zn–Mg alloy by a combination of atom probe,
  small-angle X-ray scattering and transmission electron microscopy},'' {\em
  Acta Materialia}, vol.~53, pp.~2881--2892, jun 2005.

\bibitem{Zhao2018}
H.~Zhao, F.~{De Geuser}, A.~{Kwiatkowski da Silva}, A.~Szczepaniak, B.~Gault,
  D.~Ponge, and D.~Raabe, ``{Segregation assisted grain boundary precipitation
  in a model Al-Zn-Mg-Cu alloy},'' {\em Acta Materialia}, vol.~156,
  pp.~318--329, sep 2018.

\bibitem{Mouton2019}
I.~Mouton, A.~J. Breen, S.~Wang, Y.~Chang, A.~Szczepaniak, P.~Kontis, L.~T.
  Stephenson, D.~Raabe, M.~Herbig, T.~B. Britton, and B.~Gault,
  ``{Quantification Challenges for Atom Probe Tomography of Hydrogen and
  Deuterium in Zircaloy-4},'' {\em Microscopy and Microanalysis}, vol.~25,
  no.~2, pp.~481--488, 2019.

\bibitem{lee_han_2015}
Y.-K. Lee and J.~Han, ``Current opinion in medium manganese steel,'' {\em
  Materials Science and Technology}, vol.~31, no.~7, p.~843–856, 2015.

\bibitem{kuzmina_herbig_ponge_sandlobes_raabe_2015}
M.~Kuzmina, M.~Herbig, D.~Ponge, S.~Sandlobes, and D.~Raabe, ``Linear
  complexions: Confined chemical and structural states at dislocations,'' {\em
  Science}, vol.~349, no.~6252, p.~1080–1083, 2015.

\bibitem{kuzmina_ponge_raabe_2015}
M.~Kuzmina, D.~Ponge, and D.~Raabe, ``Grain boundary segregation engineering
  and austenite reversion turn embrittlement into toughness: Example of a 9
  wt.\% medium mn steel,'' {\em Acta Materialia}, vol.~86, p.~182–192, 2015.

\bibitem{KwiatkowskidaSilva2018}
A.~K. da~Silva, D.~Ponge, Z.~Peng, G.~Inden, Y.~Lu, A.~Breen, B.~Gault, and
  D.~Raabe, ``Phase nucleation through confined spinodal fluctuations at
  crystal defects evidenced in fe-mn alloys,'' {\em Nature Communications},
  vol.~9, Mar. 2018.

\bibitem{KWIATKOWSKIDASILVA2019109}
A.~K. da~Silva, R.~D. Kamachali, D.~Ponge, B.~Gault, J.~Neugebauer, and
  D.~Raabe, ``Thermodynamics of grain boundary segregation, interfacial
  spinodal and their relevance for nucleation during solid-solid phase
  transitions,'' {\em Acta Materialia}, vol.~168, pp.~109 -- 120, 2019.

\bibitem{Maury1993}
C.~Maury, L.~Rabenberg, and C.~H. Allibert, ``Genesis of the cell
  microstructure in the sm(co, fe, cu, zr) permanent magnets with 2:17 type,''
  {\em physica status solidi (a)}, vol.~140, no.~1, pp.~57--72, 1993.

\bibitem{Gutfleisch2011}
O.~Gutfleisch, M.~A. Willard, E.~Br{\"{u}}ck, C.~H. Chen, S.~G. Sankar, and
  J.~P. Liu, ``{Magnetic Materials and Devices for the 21st Century: Stronger,
  Lighter, and More Energy Efficient},'' {\em Advanced Materials}, vol.~23,
  pp.~821--842, feb 2011.

\bibitem{Duerrschnabel}
M.~Duerrschnabel, M.~Yi, K.~Uestuener, M.~Liesegang, M.~Katter, H.~J. Kleebe,
  B.~Xu, O.~Gutfleisch, and L.~Molina-Luna, ``Atomic structure and domain wall
  pinning in samarium-cobalt-based permanent magnets,'' {\em Nature Comm.},
  vol.~8, no.~54, pp.~1--7, 2017.

\bibitem{Scheer2011}
H.~S. R.~Scheer, {\em Chalcogenide Photovoltaics: Physics, Technologies, and
  Thin Film Devices}.
\newblock Weinheim: Wiley-VCH Verlag GmbH, 2011.

\bibitem{DeVos1980}
A.~{De Vos}, ``{Detailed balance limit of the efficiency of tandem solar
  cells},'' {\em Journal of Physics D: Applied Physics}, vol.~13, no.~5,
  pp.~839--846, 1980.

\bibitem{Lomu2019}
A.~Lomuscio, T.~Rödel, T.~Schwarz, B.~Gault, M.~Melchiorre, D.~Raabe, and
  S.~Siebentritt, ``{Quasifermi-level splitting of Cu-poor and Cu-rich CuInS2
  absorber layers},'' {\em Physical Review Applied}, vol.~11, 05 2019.

\bibitem{SCHWARZ2020105081}
T.~Schwarz, A.~Lomuscio, S.~Siebentritt, and B.~Gault, ``{On the chemistry of
  grain boundaries in CuInS2 films},'' {\em Nano Energy}, vol.~76, p.~105081,
  2020.

\bibitem{Liebl1967}
H.~Liebl, ``{Ion microprobe mass analyzer},'' {\em Journal of Applied Physics},
  vol.~38, no.~13, pp.~5277--5283, 1967.

\bibitem{WITTMAACK197539}
K.~Wittmaack, ``Pre-equilibrium variation of the secondary ion yield,'' {\em
  International Journal of Mass Spectrometry and Ion Physics}, vol.~17, no.~1,
  pp.~39 -- 50, 1975.

\bibitem{Magee1978}
C.~W. Magee, W.~L. Harrington, and R.~E. Honig, ``{Secondary ion quadrupole
  mass spectrometer for depth profiling - Design and performance evaluation},''
  {\em Review of Scientific Instruments}, vol.~49, no.~4, pp.~477--485, 1978.

\bibitem{Yang2020}
L.~Yang, R.~Choi, Y.~Zheng, M.~H.~S. Bidabadi, A.~Rehman, C.~Zhang, H.~Chen,
  and Z.-G. Yang, ``{Spalling resistance of thermally grown oxide based on
  NiCoCrAlY(Ti) with different oxide peg sizes},'' {\em Rare Metals}, 2020.

\bibitem{Reddy1977}
D.~R. Reddy, ``Speech understanding systems: A summary of results of the
  five-year research effort.,'' 1977.

\bibitem{LightBGM}
Microsoft, ``Lightgbm.'' \url{https://github.com/microsoft/LightGBM}, 2017.

\end{thebibliography}

\end{document}